\documentclass[twocolumn,prx,superscriptaddress,floatfix]{revtex4-1}
\usepackage{multirow}
\usepackage{booktabs}
\usepackage{amsmath,amssymb,mathrsfs,amsthm}
\usepackage{natbib}
\usepackage{tabularx}
\usepackage{amsfonts}
\usepackage{subcaption}
\usepackage{amsmath}
\usepackage{comment}
\usepackage{bbold} 
\usepackage{hhline}
\usepackage{braket}
\usepackage{txfonts}
\usepackage{balance}
\usepackage{physics}
\usepackage{graphicx}
\usepackage{dcolumn}
\usepackage{bm}
\usepackage[utf8]{inputenc}
\usepackage[english]{babel}
\usepackage[T1]{fontenc}
\usepackage{mathtools}
\usepackage[dvipsnames]{xcolor}
\usepackage[unicode=true,bookmarks=true,bookmarksnumbered=false,bookmarksopen=false,breaklinks=false,pdfborder={0 0 1},backref=false,colorlinks=true,citecolor=blue,linkcolor=blue]{hyperref}

\def\be{\begin{equation}}
\def\ee{\end{equation}}
\def\bea{\begin{eqnarray}}
\def\eea{\end{eqnarray}}

\newcommand{\RNum}[1]{\uppercase\expandafter{\romannumeral #1\relax}}

\begin{document}

\title{Resource and entanglement study of a hybrid qudit-qubit quantum algorithm for solving the integer programming problem}% Force line breaks with \\

\author{Kapil Goswami} 
\email{kapil.goswami@uni-hamburg.de}
\affiliation{%
 Zentrum f\"ur Optische Quantentechnologien, Universit\"at Hamburg, Luruper Chaussee 149, 22761 Hamburg, Germany
}%
\author{Rick Mukherjee}%
\email{rick-mukherjee@utc.edu}
\affiliation{%
 Department of Physics \& Astronomy, University of Tennessee, Chattanooga, TN 37403, USA
}%

\affiliation{%
 UTC Quantum Center, University of Tennessee, Chattanooga, TN 37403, US
}%
\author{Peter Schmelcher}
\email{peter.schmelcher@uni-hamburg.de}
\affiliation{%
 Zentrum f\"ur Optische Quantentechnologien, Universit\"at Hamburg, Luruper Chaussee 149, 22761 Hamburg, Germany
}%

\date{\today}

\begin{abstract}
Recently, a hybrid qudit-qubit algorithm \cite{myqudit} was presented for solving the integer programming problem with a polynomial quantum advantage. In this work, we investigate the algorithm \cite{myqudit} to understand the role of qudits ($d$-dimensional quantum system) by conducting a comparative resource analysis with its qubit-only implementation and the classical simulability of the algorithm by exploring the entanglement structure. The resource analysis part is performed in terms of logical gate counts, and fault-tolerant physical resources, including non-Clifford gates. The hybrid qudit-qubit implementation has a more compact structure of the unitary operators as opposed to its qubit-only simulation which reduces the logical and fault-tolerant resource requirements. Two example problems, one with qutrits ($d=3$) and the other with ququint ($d=5$), when contrasted with their qubit-only implementation, within a simplified fault-tolerant resource model, showed $\sim 180 \times $ and $\sim 2220\times$ fewer total resource count, respectively. For the entanglement study, volume-law-like entropy growth and signatures of multi-partite entanglement is observed leading to increasing difficulty in the classical simulation of the algorithm. The algorithm generates synergistic tri-partite entanglement even for a quadratic problem, revealing that the algorithmic structure itself can generate higher-order entanglement in the system independent of the problem.     
\end{abstract}
\maketitle

\section{Introduction}

The development of quantum algorithms is motivated by solving problems that are intractable for classical computers \cite{dalzell2023quantum,montanaro2016quantum,cerezo2021variational,shor2002introduction}.
Notably, Shor showed an exponential advantage for prime factorization \cite{shor1994algorithms} and Grover established a quadratic advantage for unstructured search \cite{grover1997quantum}. Similarly, \cite{myqudit} presented a hybrid qudit-qubit quantum algorithm with quadratic speedup for polynomial integer programming problem. This leads to two fundamental questions, (i) why explore qudit-based quantum computing? \cite{wang2020qudits}  and (ii) can this algorithm be simulated classically? \cite{pan2022simulation,orus2019tensor} Both of these questions are addressed in this work by taking the hybrid algorithm \cite{myqudit} as a testbed for the analysis.

Traditionally quantum computers have used qubits, two-level quantum systems, for information processing. 
However, the scalability and expressivity issues of the current hardware present opportunities to go beyond qubits (and different qubit encoding \cite{goswami2026solving}) and utilize qudits, \(d\)-level systems as an alternative \cite{Kiktenko2023,brennen2005criteria,wills2026review,jankovic2024noisy,goswami2024integer}. By expanding the local Hilbert space to \(\mathbb{C}^d\), qudits offer denser information encoding, and richer entanglement structures \cite{burshtein2026robust,ringbauer2022universal,omanakuttan2023qudit}. Moreover, multi-particle entangling gates can be performed more compactly on qudits, often without the need for ancillas \cite{shi2026efficient,Kiktenko2023,Nikolaeva2024}. Recently, there have been many developments in quantum computing using qudits \cite{wang2020qudits,Chi2022,Kiktenko2023,Deller2023,Nikolaeva2024,Kim2024,Pudda2024,goswami2024integer,myqudit}. These include qudit-based variational approaches for solving combinatorial optimization problems, such as qudit-QAOA to formulate integer optimization problems including graph coloring and electric-vehicle charging \cite{Deller2023}, and qutrit QAOA for graph 3-coloring \cite{Bottrill2023}. Hence, we study a hybrid qudit-qubit based architecture, especially in the fault-tolerant regime and contrast it against the traditional qubit-based quantum computing.
In this work, a detailed resource analysis is conducted, quantifying the overhead of using qubit-only implementation of the hybrid qudit-qubit algorithm. This is done by counting the number of logical gates in both the implementations, which shows that the hybrid approach is more efficient ($\sim 7 \text{ to } 8$ times for qutrits and ququints). Furthermore, the study includes the resources needed for fault-tolerant computing, specifically the non-Clifford gates and the physical resources needed for each of the gates \cite{bravyi2005universal}. Using the current error correction schemes for surface codes and magic state distillation along with a simplified fault-tolerant resource model, we show that even qutrits and ququints are more efficient than qubits ($184\times$ and $2220\times$ cheaper, respectively, for the example problem), with the potential for the advantage to grow further with prime-dimensional qudits \cite{howard2012qudit,anwar2012qutrit,campbell2014enhanced,prakash2025low,PhysRevLett.123.070507,p7c9-x1m9,haah2017magic}. Hence from the efficiency perspective, qudit-based architecture can potentially be more feasible in the near-term for quantum computation.

To establish whether a quantum algorithm is classically simulable or not, one needs to study the source of quantum advantage. 
This is because to demonstrate a genuine quantum advantage, the quantum algorithms need to at least surpass classical simulation methods, such as tensor network based methods \cite{pan2022simulation,orus2019tensor}. The notion of ``quantumness'' in the system is generally used as a tool to identify the source of quantum advantage. This refers to non-classical features that distinguish quantum from classical computation \cite{aharonov1999quantum,adesso2016measures}.
One of the widely studied ways to study ``quantumness'' is entanglement, which can be a resource for quantum speedup and is often used to characterize near-term and early fault-tolerant algorithms \cite{yu2021advancements,ding2007review,bharti2022noisy,campbell2017roads}. 
It is shown that the quantum circuits with bounded von-Neumann entropy (area-law-like) lead to their efficient classical simulation via tensor-network methods \cite{berezutskii2025tensor,latorre2009short}, while circuits that generate volume-law and multipartite synergistic entanglement can become classically expensive to simulate using tensor-network methods \cite{PhysRevLett.71.1291,amico2008entanglement,cerf1998information,jozsa2006simulation,m2019tripartite,seshadri2018tripartite,vedral2002role,caceffo2023negative}. In this work, we utilize these entanglement indicators to show that the hybrid qudit-qubit algorithm is classically difficult to simulate.
The generation of the entanglement (von Neumann entropy) is studied analytically between the qudit-qubit subsystems of the hybrid algorithm for the intermediate states and validated using numerical simulation. This showed a correlation between the complex part of a particular problem and high entanglement generation in the part of the circuit tackling that complexity. The difficulty of classical simulability of the algorithm is then shown by numerically analyzing the entanglement growth with the system size for each state (volume law-like behavior) coupled with multi-partite (Bi-, tri-, and 4-partite) mutual information as indicators for global entanglement generation.

The manuscript is as follows. Sec.~\ref{Overview} presents an overview of the hybrid quantum algorithm.  Section~\ref{quditqubit} develops a resource-theoretic comparison between the qudit-native and qubit-encoded implementations, using logical gate counts, and the fault-tolerant physical resource cost including magic-state distillation comparison. Sec.~\ref{Entanalg} discusses the generation of von-Neumann entropy between the qudit-qubit register, and the entanglement growth with bipartite/tripartite/four-partite mutual information to establish volume-law, synergistic multipartite entanglement, all indicating towards inefficient classical simulation. Finally, Section~\ref{discuss} provides our conclusions and discusses implication of the two results for near-term quantum hardware.

\section{Overview of the hybrid qudit-qubit algorithm}
\label{Overview}

\begin{figure*}[t]
    \centering
    \includegraphics[width = 1\linewidth]{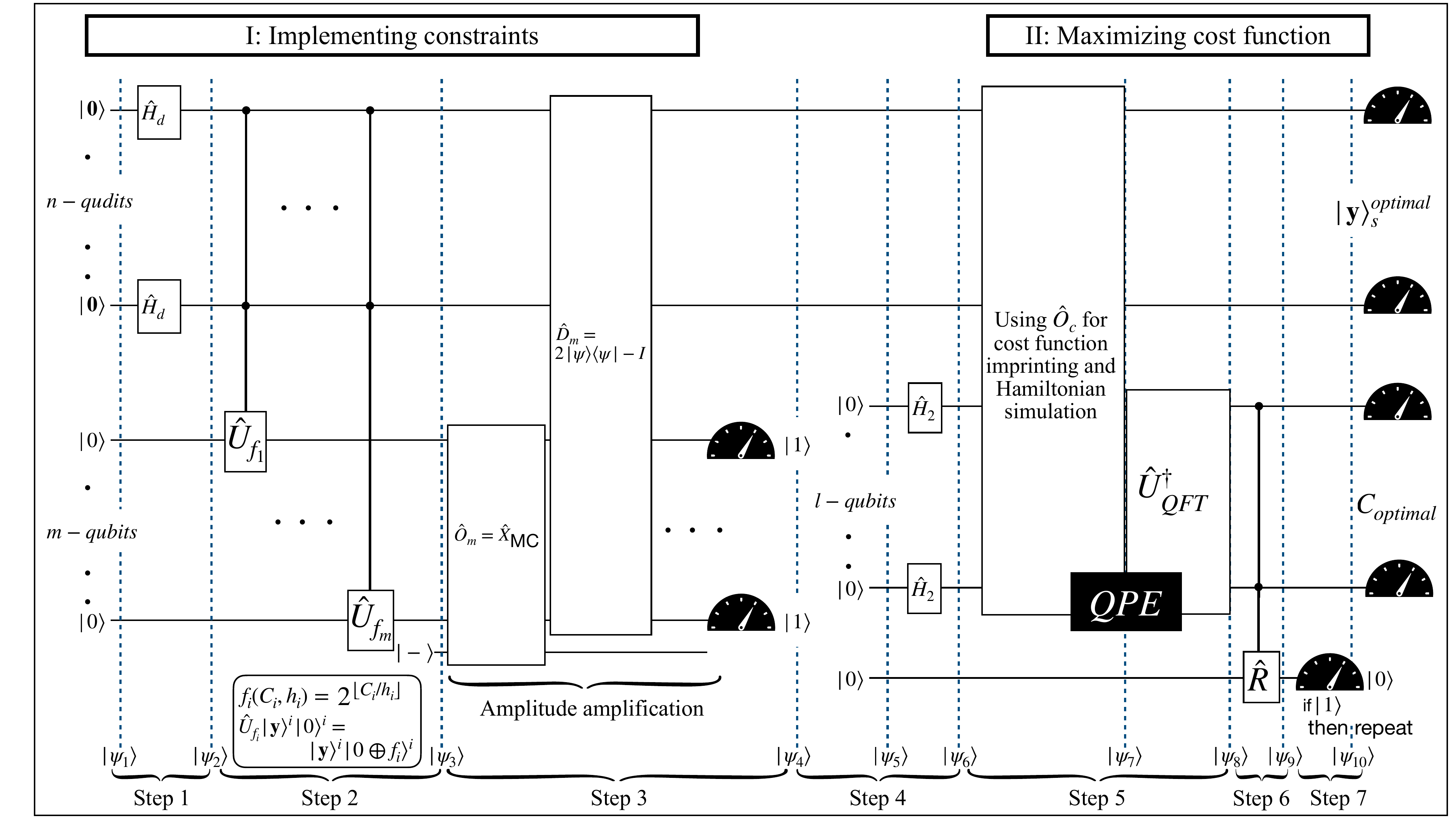}
    \caption{The schematic circuit for the quantum algorithm to solve a general IP problem with $n$ variables and $m$ constraints. The stage I implements all the constraints, the stage II maximizes the cost function and the measurements at the end are probabilistic. The $n$-qudit register stores the integer variables, and the $m$-qubit register indicates the satisfiability of the constraints. The subroutines for the constraints satisfying stage include $n$-Hadamard gates $\hat{H}_d$ for qudits, $m$ one-sparse entangling unitary operators $\hat{U}_{f_i}$ in sequence, $\sim d^{n/2}$ times application of the Grover operator $\hat{O}_m\cdot \hat{D}_m $ for amplitude amplification, and $m$-measurements for the qubit register. The subroutines for the next part are: $l$-Hadamard gates $\hat{H}_2$ for qubits, cost function imprinting using a QPE procedure by Hamiltonian simulation of $\hat{O}_c$, a multi-qubit controlled rotation gate $\hat{R}$, and $l+1$-qubit plus $n$-qudit measurements. Taken from Ref.~\cite{myqudit}.}
    \label{Dalg}
\end{figure*}

Integer programming (IP) is a ubiquitous but NP-hard combinatorial optimization problem whose complexity increases exponentially with problem size and is suited for a direct encoding with qudits \cite{wolsey2020integer,schrijver1998theory,goswami2024integer}. The quantum algorithm presented in Ref.~\cite{myqudit} solves the integer programming problem using a hybrid qudit-qubit implementation. To the set the stage and be self-contained we briefly summarize here the hybrid qudit-qubit algorithm. The hybrid algorithm solves the problem using a circuit consisting of two stages, the schematic circuit diagram of the algorithm is shown in Fig.~\ref{Dalg} which is taken from Ref.~\cite{myqudit}.
Stage I addresses the constraint satisfaction (step 1 to step 3) and stage II addresses the cost function optimization (step 4 to step 7).:
\begin{itemize}
    \item Step 1. Initialization: The system is initialized with a data register of $n$ qudits and a constraint register of $m$ qubits. $n$-Hadamard gates are applied to each of the $n$ qudit register to create an equal superposition of all the possible states.
    \item Step 2. Implementation of the constraint-satisfying distillation function: The unitary operator $\hat{U}_{f_i}$ is applied for each constraint $i$. $\hat{U}_{f_i}$ flips a representative qubit (from $\ket{0}$ to $\ket{1}$) in the constraint register if a multi-qudit state satisfies a constraint. For $m$ constraints, there are $m$ such unitary operators. This entangles the multi-qudit states satisfying all the constraints to the multi-qubit state $\ket{1}^{\otimes m}$.
    \item Step 3. Reductions to the feasible subspace of IP: This is done by applying an amplitude amplification step to the target multi-qubit state $\ket{1}^{\otimes m}$ \cite{grover1997quantum}. Subsequently, measuring the qubit register in $\ket{1}^{\otimes m}$, collapses the data register to the feasible subspace of the given IP problem instance. 
    \item Step 4. Mid-circuit initialization: The $m$-qubit register is measured (and discarded) at the end of the stage I and is replaced with $l+1$ qubits that are used in stage II.
    \item Step 5. Encoding the cost function using quantum phase estimation: A Hamiltonian simulation based \textit{quantum phase estimation} block is used to encode the cost function as the phase of each state in the feasible region. This requires a phase imprinting operator $\hat{O}_c$, that extracts the cost function value from the multi-qudit state and encode it as the phase. This uses the $l$-qubit register for quantum phase estimation.
    \item Step 6. Amplitude encoding of the cost function: A Harrow–Hassidim–Lloyd (HHL) algorithm\cite{harrow2009quantum}-like \textit{multi-controlled single-qubit rotation} is used that encodes the phase of each state to its amplitude, which in turn maximizes the probability of the optimal solution.
    \item Step 7. Final measurement for the optimal solution: The state obtained after the single qubit measurement has the highest probability of measurement for the optimal solution.
\end{itemize}
The algorithm has a quadratic quantum speed-up and has the best scaling for the non-linear polynomial IP. Although the quantum advantage is established, the algorithm can be encoded to a qubit-based architecture by using ancillary qubits. The motivation for using qudits still needs to be justified by quantifying the resource advantage between the hybrid and qubit-encoded implementation of the algorithm, as is done in the next section.
We define an exemplar non-linear IP problems and the hybrid algorithm is analyzed in the course of solving it:
\begin{equation}
\begin{split}
& \quad \quad  \quad \quad  \quad  \quad \text{Example problem}\\ 
C(\mathbf{x}) &= -x_1 - x_2 - x_3 - x_4 \\
&\quad + 2x_1x_2 + 2x_2x_3 + 2x_3x_4 + x_1x_3 + x_1x_4 + x_2x_4 \\
&\quad + x_1x_2x_3 + x_2x_3x_4 - x_1x_3x_4 + x_1x_2x_3x_4 \\
\text{To } &\text{be maximized under:} \\
C_1&: -x_1 - x_2 + 2x_1x_2 + x_1x_2x_3 \le 3 \\
C_2&: -x_3 - x_4 + 2x_2x_3 + 2x_3x_4 + x_2x_3x_4 \le 3 \\
C_3&: -x_1x_3x_4 + x_1x_3 \le 3 \\
C_4&: x_1x_2x_3x_4 + x_1x_4 + x_2x_4 \le 4 \\
x_i &\in \{0, 1, \dots, d\}.
\end{split}
\label{ExP8}
\end{equation}
In the above problem, the variables $x_i$ take values from $0$ to $d$, which in this work will lie in $\{2,3,4,5\}$.
The parameters of the algorithm considered for this study are as follows,
\begin{itemize}
    \item $n$: number of qudits (variable register), dimension $d$
    \item $k = \lceil\log_2 d\rceil$: qubits needed to encode one qudit
    \item $m$: number of constraint ancilla qubits
    \item $l$: number of QPE estimation qubits
    \item $M$: number of monomials in cost function; $r_i$: order of monomial $i$
\end{itemize}

\section{Resource analysis of the hybrid qudit-qubit algorithm and its qubit-only implementation}
\label{quditqubit}
As the field of qudit control is less mature than qubit control, hence, a quantitative resource argument is required to make a case for the inherent practical advantage of the hybrid qudit-qubit algorithm over the qubit-only implementation of the algorithm. This is argued in two steps: logical gate counts, and the fault-tolerant physical resource count using magic-state distillation.
The hybrid algorithm is translated to qubit algorithm where one qudit is replaced by $k$ qubits, increasing the initial resources by a factor $k$. Furthermore, the multi-body qudit and hybrid qubit-qudit operations are simulated via qubit operations which introduces more qubit gates. The resource overhead for converting the algorithm to all qubit encoding is quantified here. 

A $d$-dimensional qudit is encoded into $k = \lceil \log_2 d \rceil$ qubits via
\begin{equation}
V: \mathbb{C}^d \to (\mathbb{C}^2)^{\otimes k}, \qquad \ket{x}_d \mapsto \ket{b_{k-1}(x)}_2 \otimes \cdots \otimes \ket{b_0(x)}_2,
\label{eq:encoding_isometry_2}
\end{equation}
where $x = \sum_{j=0}^{k-1} b_j(x) 2^j$ is the binary expansion. Hence, the logical unit (qubit or qudit) count becomes,
\begin{align}
N_{\text{qd}} &= n + m + l \quad \text{(qudit-native)}, \label{eq:carriers_qd}\\
N_{\text{qb}} &= nk + m + l \quad \text{(qubit-encoded)}. \label{eq:carriers_qb}
\end{align}

The complexity of the resulting qubit gates is compared with the gates in the hybrid algorithm, counting the Clifford and non-Clifford gates specifically. This counting helps to estimate the physical resources required to construct each gate and to discuss the fault-tolerance resource model. 

\noindent \textbf{Importance of the non-Clifford gate overhead}:
Clifford gates are the ones that map Pauli operators to Pauli operators under conjugation \cite{grier2022classification,bravyi2005universal}. For qubits, this is the group generated by $H$, $S$, and $CNOT$ while the qudit analogues built from the generalized Hadamard, phase, and SUM gates. By the Gottesman-Knill theorem, circuits built entirely from Clifford gates are efficiently simulable on a classical computer, admitting efficient, often transversal, implementation in most quantum error-correcting codes \cite{gottesman1998heisenberg}. Non-Clifford gates (e.g. the qubit $T$ gate) provides a circuit universality, but they generally cannot be implemented
transversally \cite{campbell2017roads}. Instead they require magic-state injection and distillation techniques \cite{liu2023magic,sales2025experimental,campbell2012magic}, leading to additional physical qubits, repeated error-correction cycles, and more resource overhead to prepare high-fidelity states. Preparing a fault-tolerant non-Clifford gate is more expensive than a Clifford gate, hence the count of the two gate types determines the real hardware cost \cite{kissinger2020reducing,kobayashi2026clifford}.

This is one of the important factors while comparing the hybrid algorithm with its qubit-only implementation. For instance, the qubit-based $\hat{O}_c$ operator requires $O(r^2k^2)$ non-Clifford gates, while the hybrid version needs only $O(r)$, this is further shown in detail later in this section for all the operations in the algorithm. The qubit disadvantage is due to the larger total gate count and the high cost of its dominant non-Clifford component. The hybrid implementation reduces the total number of logical operations, thus, lowering the overall error-correction overhead. It has a larger fraction of Clifford operations which reduces the need for magic-state distillation. The fault-tolerant resource advantage can thus far exceed the usual logical gate counting. 

\subsection{Logical gate complexity estimates for each part of the algorithm}

In this section, each gate in the qudit-qubit hybrid algorithm is decomposed into qubit gates. We take into account an all to all connected as well as nearest neighbor architecture to count the number of gates. This decomposition is done for each stage of the algorithm, and a comparative analysis is conducted between the qubit encoded algorithm and the original hybrid algorithm. To this end, we identify the major operations in each stage that require careful resource count, $\hat{U}_{f_i}$ and amplitude amplification in Stage~I whereas $\hat{O}_c$ and QPE in Stage~II. The total gate count as well as the non-Clifford gate count are also presented here. Appendix~\ref{cliffnocliff} contains the table which categorizes the typical gates as Clifford and non-Clifford for both the hybrid and qubit implementations. The gate counts derived below are leading-order resource estimates and depend on the specific method used to decompose the operators into elementary gates.  

\textbf{Constraint Unitaries $\hat{U}_f$:} The constraint unitaries are $1$-sparse by construction. Every computational basis state maps to exactly one other basis state, possibly with a phase. This is essentially a classical reversible computation \cite{bennett1997strengths}. The gate count estimates used in this resource analysis are derived via an explicit multi-controlled-gate decomposition.

\textit{Qudit-qubit implementation.} Any permutation can be decomposed into a series of transpositions that swap two elements. On all-to-all connected hardware, a
permutation on $n$ elements requires at most $n-1$ swaps between arbitrary pairs. Each transposition between two qudit states is implemented using a generalized controlled-SWAP gate.
To implement a transposition between states $a$ and $b$, it is needed to be checked if the qudit is in state $a$ or $b$ which requires multi-controlled operations. For example, to swap amplitudes of states $3$ and $7$ in a single qudit of dimension $8$, a rotation is applied in the two-dimensional subspace spanned by those states. This can be done by first applying single-qudit gates to map states $3$ and $7$ to states $0$ and $1$, performing a SWAP in that subspace (which is just an $X-$gate), and mapping back.
The Clifford gates typically arise from the mapping operations, these are generalized $X-$gates (Clifford for qudits and qubits both) and SUM gates used in controlled operations. The non-Clifford gates come from the phase corrections needed, when decomposing a permutation into transpositions. Each transposition need a controlled-Z phase gate to fix relative phases. So for $n$ transpositions, approximately $2n$ SUM gates (Clifford) and $n$ phase gates (non-Clifford) are needed. Hence, hybrid implementation of $\hat{U}_f$ is about $67\%$ Clifford and $33\%$ non-Clifford operations.
On all-to-all connected hardware,
\begin{equation}
G_{U_f}^{qd,\mathrm{AA}} = m\,(n-1)\cdot g_{\text{transposition}},
\label{eq:uf_qd_aa}
\end{equation}
where $g_{\text{transposition}}$ is the cost per qudit transposition, typically 1-3 two-qudit SUM gates. While in the nearest-neighbor connectivity, transpositions can only act on adjacent qudits. An arbitrary permutation requires up to $\binom{n}{2}$ adjacent transpositions, so each multi-controlled permutation on $n$ qudits decomposes as
\begin{equation}
G_{U_f}^{qd,\mathrm{NN}} = m\binom{n}{2}\cdot g_{\text{transposition}}.
\label{eq:uf_qd_nn}
\end{equation}
\begin{table*}[t]
\centering
\renewcommand{\arraystretch}{1.3}
\begin{tabular}{|l|c|c|c|c|c|c|}
\hline
\multirow{2}{*}{Stage} & \multicolumn{2}{c|}{All-to-all} & \multicolumn{2}{c|}{Nearest neighbor} & \multicolumn{2}{c|}{Non-Clifford ratio} \\
\cline{2-7}
 & Qudit & Qubit & Qudit & Qubit & Qudit & Qubit \\
\hline
$\hat{U}_f$ (constraint checks) & $\mathcal O(n)$ & $\mathcal O(n(\log_2 d)^2)$ & $\mathcal O(n^2)$ & $\mathcal O(n^2(\log_2 d)^4)$ & $33\%$ & $47\%$ \\
\hline
Amplitude amplification & $\mathcal O(d^{n/2}\!\cdot\! n)$ & $\mathcal O(d^{n/2}\!\cdot\! n(\log_2 d)^3)$ & $\mathcal O(d^{n/2}\!\cdot\! n^2)$ & $\mathcal O(d^{n/2}\!\cdot\! n^2(\log_2 d)^4)$ & $29\%$ & $46\%$ \\
\hline
$\hat O_c$ (state preparation) & $\mathcal O(M\bar r)$ & $\mathcal O(M\bar r^2 k^2)$ & $\mathcal O(M\bar r\, n)$ & $\mathcal O(M\bar r^2 k^2 n)$ & $100\%$ & $100\%$ \\
\hline
QPE$^{\ast}$ (controlled-$\hat O_c$ + QFT) & $\mathcal O((2^l{-}1)M\bar r + l^2)$ & $\mathcal O((2^l{-}1)M\bar r^2 k^2 + l^2)$ & $\mathcal O((2^l{-}1)M\bar r\, n + l^2)$ & $\mathcal O((2^l{-}1)M\bar r^2 k^2 n + l^2)$ & $99\%$ & $100\%$ \\
\hline
\end{tabular}
\caption{Leading-order gate-count scaling for the four dominant operations in the hybrid algorithm, under all-to-all and nearest-neighbor connectivity, together with each operation's non-Clifford gate fraction (see Sec.~III.A for derivations). $M$ is the number of monomials in the cost/constraint function with $r_i$ being order of the monomial $i$ and $\bar r$ their average degree;  $n$ is the number of qudits (variable register) of dimension $d$, $m$ is the number of constraint ancilla qubits, and $l$ is the number of QPE estimation qubits. Qubit costs are for the encoded implementation of the same operation. $^{\ast}$The QPE stage includes $2^l-1$ applications of the controlled cost operator plus the inverse QFT.}
\label{tab:gate-complexity-summary}
\end{table*}

\textit{Qubit implementation.} The same permutation becomes more complex because each qudit is encoded into $k$ qubits, the permutation now acts on $nk$ qubits total. Any classical reversible function on $nk$ bits can be built from Toffoli gates (controlled-controlled-NOT) and NOT gates. A Toffoli gate decomposes into exactly 6 CNOT gates, 7 $T$ gates, and 2 Hadamard gates \cite{shende2008cnot}. The CNOTs and Hadamards are Clifford while the 7 $T$ gates are non-Clifford. So each Toffoli gives 8 Clifford operations and 7 non-Clifford operations, leading to $53\%$ Clifford and $47\%$ non-Clifford gates.
Unlike the hybrid case, where a single operation tests the value of a qudit directly, in the qubit case, it has to be a $k$-qubit block to match a target $k$-qubit block. This amounts to $k$-controlled operations costing $O(\log_2{d})$ Toffolis per qudit ($k$-qubits). For a permutation on $nk$ bits, this brings the typical Toffoli count to $O(nk\log_2{d})$ using standard synthesis methods. Each Toffoli contributes 7 non-Clifford gates, hence the absolute count of non-Clifford gates is much higher than the hybrid version. In the qubit encoding, the same operation becomes a multi-controlled gate
on $nk$ controls, each carrying this additional per-slot decoding overhead.
On all-to-all connected hardware,
\begin{equation}
G_{\hat U_f}^{\text{qb},\mathrm{AA}} = m \cdot \mathcal{O}(n(\log_2{d})^2)\cdot g_{\text{Toffoli}},
\label{eq:gates_uf_qb}
\end{equation}
where $g_{\text{Toffoli}}$ is the count of Toffoli gates needed. And on nearest-neighbor hardware, where the same adjacent-only routing penalty applies as in the qudit case (Eq.~\eqref{eq:uf_qd_nn}), it is,
\begin{equation}
G_{\hat U_f}^{\text{qb},\mathrm{NN}} = m \cdot \mathcal{O}((n(\log_2{d})^2)^2)\cdot g_{\text{Toffoli}}.
\label{eq:gates_uf_qb2}
\end{equation}

\noindent \textbf{Decomposing Amplitude Amplification}:
The amplitude amplification stage consists of the Grover diffusion operator applied iteratively. The state of the algorithm prior to amplification is $|\psi_3\rangle = \hat U_f H^{\otimes n}|0\rangle$ rather than the equal superposition, the diffusion operator must reflect about $|\psi_3\rangle$, not about $H^{\otimes n}|0\rangle$. Each iteration applies $A = \hat U_f H^{\otimes n}$, a multi-controlled phase flip about $|0\rangle$, and then $A^\dagger = (H^{\otimes n})^\dagger \hat U_f^\dagger$. The diffusion operator has three parts: the Hadamard layers, the constraint unitary $\hat U_f,\hat U_f^\dagger$, and the multi-controlled phase flip.

\textit{Qudit-qubit implementation.} The multi-controlled phase flip decomposes into a generalized Toffoli-like structure with one non-Clifford phase gate and $O(n)$ Clifford SUM gates. The Hadamard layers are single-qudit Fourier transforms $H_d$; for prime qudit dimension $d$, $H_d$ and its inverse are Clifford gates, requiring no non-Clifford rotations. The two applications of $\hat U_f$, however, contribute to the non-Clifford count as discussed above ($\approx 30\%$ of $G_{U_f}^{\text{qd}}$ each). Each Grover iteration contributes $1 + 2\times(\text{non-Clifford count of }\hat U_f)$ non-Clifford gates. Since $\hat U_f$ and the phase flip are the same leading order in $n$, the non-Clifford fraction per iteration converges to close to that of $\hat U_f$ itself ($\approx 29\%$, Table~\ref{tab:gate-complexity-summary}).

\textit{Qubit implementation.} The multi-controlled phase flip with $nk$ controls is expensive. The standard decomposition without ancilla requires $O(nk)$ Toffoli gates, each contributing 7 non-Clifford $T$ gates. With ancilla qubits, this can be reduced to $O(nk)$ total gates but the non-Clifford count remains proportional. The Hadamard layers become single-qubit Hadamards, which are Clifford. As in the qudit case, the two per-iteration applications of $\hat U_f$ dominate the non-Clifford budget, since $G_{U_f}^{\text{qb}}$ is at least as large as the phase-flip cost $O(nk)$; the resulting non-Clifford fraction per iteration is close to the $\hat U_f$ cost ($\approx 46\%$). The qubit-vs-qudit gap in absolute non-Clifford count is therefore set primarily by the $\hat U_f$ gap which is already established above, rather than by the phase-flip decomposition.

\noindent \textbf{Cost Operator $\hat{O}_c$:} 
The cost operator $\hat{O}_c$ applies a phase to each basis state based on the cost function $C_f$. If the cost function is a sum of $r$-body terms, each term in $\hat{O}_c$ is a product of $r$ qudit operators of the form $q_1^{p_1} q_2^{p_2} q_3^{p_3}$.

\textit{Qudit-qubit implementation.} 
An \(r\)-body diagonal term is constructed through a tree-like sequence of two-qudit operations that builds the required multi-qudit phase, which provides a leading-order gate cost \(O(r)\). 
For example, for a three-body term on qudits 1, 2, and 3, the required three-body phase is accumulated through a tree-like sequence of two-qudit phase operations. 
Each phase operator with an arbitrary phase angle is non-Clifford, and no additional Clifford gates are counted at leading order, since the entire operation is diagonal in the computational basis. 
Thus, for an \(r\)-body interaction, the non-Clifford gate count scales as \(O(r)\). 
For the simple tree-based decomposition, we take \(r-1\) two-qudit phase operations as the gate count. 
Hence, the Clifford count for \(\hat{O}_c\) is taken to be zero at leading order for the qudit implementation.
For a cost function with \(M\) monomials of body order \(r_i\), the qudit implementation therefore costs
\begin{equation}
G_{O_c}^{\text{qd}}
=
\sum_{i=1}^{M}
\max(r_i-1,0)
\qquad
\text{(two-qudit gates)}.
\label{eq:gates_oc_qd}
\end{equation} 
Using \(\bar r\) as the average order of the monomials in the cost function, this gives the leading-order scaling
\begin{equation}
G_{O_c}^{\text{qd}}
=
O(M\bar r).
\label{eq:gates_oc_qd_scaling}
\end{equation}

\textit{Qubit implementation.} 
The same \(r\)-body qudit term is encoded over \(rk\) qubits. 
Within the pairwise binary-decomposition model, the encoded interaction is represented through pairwise interactions among the \(rk\) qubits. 
The number of possible qubit pairs among \(rk\) qubits is \(\binom{rk}{2}\), which grows quadratically with \(rk\). 
Each pairwise phase interaction with an arbitrary angle is counted as a non-Clifford controlled-phase gate. 
Hence, the qubit implementation requires \(O(r^2k^2)\) non-Clifford gates compared with the \(O(r)\) leading-order cost of the corresponding qudit implementation.
Additionally, if the individual qudit powers \(p_j \geq 2\), intra-qudit contributions also appear. 
A single-qudit operator \(q^p\), when encoded into \(k\) qubits, generally produces additional terms involving multiple qubits. 
These introduce an additional scaling as \(O(k^2)\) non-Clifford controlled-phase operations for each encoded qudit term.
The qubit encoding is,
\begin{equation}
G_{O_c}^{\text{qb}}
=
\sum_{i=1}^{M}
\binom{r_i k}{2}
=
\mathcal{O}(M\bar{r}^{\,2}k^2).
\label{eq:gates_oc_qb}
\end{equation}

The operator complexity gap per application of \(\hat{O}_c\) is therefore
\begin{equation}
\mathcal{G}_{O_c}
=
\frac{G_{O_c}^{\text{qb}}}
     {G_{O_c}^{\text{qd}}}
=
\mathcal{O}(\bar{r}k^2),
\label{eq:gap_oc}
\end{equation}
which grows with both the encoding overhead \(k\) and the average monomial order \(\bar r\). The \(\hat{O}_c\) stage exhibits the largest resource gap between the two implementations within the decomposition considered here. 
The native multi-level structure of qudits help to represent the multi-body terms directly, whereas the qubit encoding distributes the same information over multiple two-level systems and therefore introduces an additional binary-encoding overhead.
The resulting overhead therefore corresponds to the decomposition model chosen for this analysis and may differ for alternative synthesis methods.

\noindent \textbf{QPE routine:}
The quantum phase estimation circuit has two main components: controlled applications of the cost operator (controlled-\(\hat{O}_c\)) and the inverse quantum Fourier transform. 
The Clifford and non-Clifford split is evaluated for the corresponding controlled operations, which introduce an additional control to \(\hat{O}_c\). 
In the decomposition model considered here, the operators \(\hat{O}_c^{\,2^j}\) are implemented through repeated applications of the controlled-\(\hat{O}_c\) operation. 
The total number of such applications is therefore
\begin{equation}
\sum_{j=0}^{l-1}2^j = 2^l-1,
\end{equation}
so the corresponding gate counts are multiplied by \(2^l-1\). 

\begin{table}[t]
\centering
\renewcommand{\arraystretch}{1.3}
\begin{tabular}{|l|c|c|c|c|c|c|}
\hline
\multirow{2}{*}{Stage} & \multicolumn{2}{c|}{All-to-all} & \multicolumn{2}{c|}{Nearest neighbor} & \multicolumn{2}{c|}{NC ratio} \\
\cline{2-7}
 & Qudit & Qubit & Qudit & Qubit & Qudit & Qubit \\
\hline
$\hat{U}_f$ & 16 & 40 & 64 & 404 & $33\%$ & $47\%$ \\
\hline
Amplitude amp. & 36 & 143 & 144 & 908 & $29\%$ & $46\%$ \\
\hline
$\hat O_c$ (state prep) & 29 & 240 & 116 & 959 & $100\%$ & $100\%$ \\
\hline
QPE & 212 & 1689 & 821 & 6722 & $99\%$ & $100\%$ \\
\hline
Total gates & 293 & 2112 & 1145 & 8993 & -- & -- \\
\hline
Gate ratio (qb/qd) & \multicolumn{2}{c|}{7.21} & \multicolumn{2}{c|}{7.85} & -- & -- \\
\hline
Total non-Clifford & 254 & 2014 & 992 & 8289 & -- & -- \\
\hline
NC ratio (qb/qd) & \multicolumn{2}{c|}{7.93} & \multicolumn{2}{c|}{8.36} & -- & -- \\
\hline
\end{tabular}
\caption{Leading-order gate-count along with the non-Clifford gate ratio for the four dominant operations in the hybrid algorithm, under all-to-all and nearest-neighbor connectivity for the worked example problem from Eq.~(\ref{ExP8}) with $n=4$, $m=4$, $d=3$, $k=2$, $l=3$, $M=14$, $\bar r\approx2.071$. The table also shows the total gate count, total non-Clifford gate count, and their ratio for comparing the resource count in both the cases.}
\label{tab:gate-complexity-l3}
\end{table}

\textit{Qudit-qubit implementation.} The QFT part for $l$ ancilla qubits requires $O(l^2)$ controlled-phase gates between the ancilla qubits. Every controlled-phase with an arbitrary angle is non-Clifford. The Hadamard gates are Clifford. So for an $l$-qubit QFT, $l$ Clifford gates (the Hadamards) and roughly $O(l^2)/2$ non-Clifford gates (the controlled-phases) are present. 

\textit{Qubit implementation.} The structure for qubits is similar to that for the hybrid case, though each qudit operation might decompose further; since the ancilla register is kept as qubits in both cases, the QFT cost is similar.

The QPE routine applies $\hat{O}_c$ a total of $2^l - 1$ times, amplifying the gate-count difference,
\begin{equation}
\mathcal{G}_{\text{QPE}} = (2^l - 1) \cdot \mathcal{G}_{O^{qb/qd}_c} + \mathcal{O}(l^2),
\label{eq:gap_qpe}
\end{equation}
which is the QFT overhead, identical for both the implementations.
Due to the repetition, the non-Clifford count dominates over the Clifford Hadamard gates.  

\noindent \textbf{Summary:} The total gate count comparison is summarized in Tab.~\ref{tab:gate-complexity-summary}. The details of the conversion from hybrid to qubit algorithm along with the gate set used for counting are given below.

\begin{enumerate}
    \item \textbf{Native gate set.} Qudit: single-qudit rotation $R(\theta,\phi)$, two-qudit SUM (generalized CNOT). Qubit: single-qubit $R_x,R_y,R_z$, two-qubit CNOT. 
    \item \textbf{Qudits into qubits.} Each $d$-level qudit $\to$ $k$ qubits. Single-qudit $\hat q^p$ becomes polynomial in $k$ bits. $r$-body qudit interaction $\to$ interactions among all $rk$ encoded bits.
    \item \textbf{Pairwise interactions.} Qudit: one $r$-body term costs $r{-}1$ two-qudit gates. Qubit: one $r$-body term costs $\binom{rk}{2}$ two-qubit gates, plus intra-qudit terms $\binom{k}{2}$ for each power $\ge 2$.
    \item \textbf{Connectivity.} For nearest-neighbor: SWAP cost: $3$ entangling gates per unit distance. Average distance $\bar d \approx n/2$ for qudits, $\bar D \approx nk/2$ for qubits.
\end{enumerate}

The Tab.~\ref{tab:gate-complexity-summary} show the gate counts along with the non-Clifford gates fraction that quantifies the high resource gates. It provides the counting for all-to-all and nearest neighbor connectivity architectures. It is evident from the counting of the logical resources that the hybrid algorithm is more efficient compared to the qubit only implementation. The counting is applied to a small example problem given by Eq.~\ref{ExP8} and the resource count is shown is Tab.~\ref{tab:gate-complexity-l3}. The stepwise gate count calculation for the example problem is given in Appendix~\ref{gatecount}. This is further enhanced next by comparing the physical resources needed for both the implementations in the fault-tolerant regime.

\subsection{Fault-tolerant resource comparison: Qubits vs. hybrid qudit-qubit}

A comparison of fault-tolerant resources between qubits and hybrid qudit-qubit schemes is a difficult task as it requires detailed models of magic state distillation factories \cite{litinski2019game} and their interplay with the data register. Two ingredients play a major role for the comparison in the fault tolerant regime. The first is the surface code that protects the quantum information. The physical qubits and qudits are grouped together in a code with a code distance $\delta$ that determines how many simultaneous physical errors are allowed that keep the logical information reliable.
The second is magic state distillation (MSD) that is indirectly used to implement non-Clifford gate. The noisy, physically-prepared non-Clifford resource states (magic states) are distilled by consuming many raw copies to produce a few high-fidelity states. We restrict the magic-state-distillation analysis to qudits of prime dimension $d$. In this case, arithmetic modulo $d$ is over the finite field $\mathbb{F}_d$, providing the algebraic structure used in prime-dimensional distillation constructions such as
Ref.~\cite{saha2026sublogarithmic}. Extending the present analysis to composite dimensions would require a separate treatment. 

For a surface code of distance $\delta$ built from $d$-dimensional qudits, the physical carrier (physical qubits or qudits used in error correction code) count per logical unit (logical qubits or qudits) is
\begin{equation}
\kappa(\delta) = c_d\,\delta^2, \qquad c_{d=2}=2, \quad c_{d>2}=3,
\end{equation}
which is given by the standard qubit surface code scaling and its qudit generalization \cite{fowler2012surface,Marks2017}. The logical error rate at distance $\delta$ is thus,
\begin{equation}
p_L(\delta) \approx A\left(\frac{p_{\text{phys}}}{p_{\text{th}}(d)}\right)^{(\delta+1)/2},
\end{equation}
with constant $A=0.1$, physical qubit/qudit error $p_{\text{phys}}=10^{-3}$, and $p_{\text{th}}(d)$ is the dimension-dependent surface-code threshold. Given a target logical error rate $p_L^{\text{target}}(d) = (1-P_{\text{success}})/N_T(d)$ set by the total non-Clifford gate count $N_T(d)$, the required code distance can be calculated as the smallest odd integer \cite{fowler2012surface} satisfying,
\begin{equation}
\delta(d) \;>\; \frac{2\ln\!\left(p_L^{\text{target}}(d)/A\right)}{\ln\!\left(p_{\text{phys}}/p_{\text{th}}(d)\right)} - 1.
\end{equation}
A distance-$\delta$ code corrects up to $(\delta-1)/2$ errors, so an even $\delta$ and $\delta-1$ correct the same number of errors, hence, a smallest odd integer satisfying the target error rate is the resource-optimal choice.

The threshold $p_{\text{th}}(d)$ depends on the dimension $d$ as shown in Ref.~\cite{Marks2017}. The surface-code memory thresholds under the HDRG decoder \cite{watson2015fast} are $p_{\text{th}}(2) = 0.093$ and $p_{\text{th}}(5) = 0.1255$. The threshold increases monotonically with $d$ and saturates near $0.155$ as $d\to\infty$. Hence, $\delta(d{=}3)$ is bounded using the two extremes between $d=2$ and $d=5$ as $p_{\text{th}}(2)=0.093 \le p_{\text{th}}(3) \le p_{\text{th}}(5)=0.1255$.

We now evaluate the surface code resource formulas using the gate counts of Table~\ref{tab:gate-complexity-l3} for the example problem (Eq.~\ref{ExP8}), all-to-all connectivity, with $n=4$, $m=4$, $l=3$.
For the qubit-only implementation encoding a $d=3$ problem $N_T^{\text{qb}}=2014$, choosing $P_{success} = 0.99$ leads to $p_L^{\text{target}} = 0.01/2014 \approx 4.97\times10^{-6}$, giving $\delta_q = 5$ using $p_{\text{th}}(2)=0.093$. For the qutrit-hybrid implementation ($N_T^{\text{qd}}=254$), $p_L^{\text{target}} = 0.01/254 \approx 3.94\times10^{-5}$. Evaluating the distance at both extremes of the $d=3$ threshold range gives $\delta(3) = 3$ using $p_{\text{th}}=0.093$, and $\delta(3) = 3$ using $p_{\text{th}}=0.1255$. Both of them give the same integer for the distance, so $\delta(3)=3$ is a robust choice of threshold within the range.
Similar analysis can be done for the example problem with $d=5$.
For the qubit-only implementation encoding a $d=5$ problem the total non-Clifford gates $N_T^{\text{qb}}=4945$, using $k=\lceil\log_2 5\rceil=3$, $p_L^{\text{target}} \approx 2.02\times10^{-6}$, again giving $\delta_q=5$. For the ququint-hybrid implementation ($N_T^{\text{qd}}=273$), $p_L^{\text{target}} \approx 3.66\times10^{-5}$, giving $\delta(5)=3$ using the directly reported $p_{\text{th}}(5)=0.1255$.

With $\kappa_{\text{qb}}=2(5)^2=50$ and $\kappa_{\text{qd}}=3(3)^2=27$ physical carriers per logical unit for the cases $d=3$ and $d=5$, and logical unit counts $N^{\text{qb}}=nk+m+l$, $N^{\text{qd}}=n\text{ qudits}+(m{+}l)\text{ qubits}$ evaluated at $l=3$,
\begin{align}
Q_{\text{register}}^{\text{qb}}(d{=}3\text{ instance}) &= 15\times50 = 750, \\
Q_{\text{register}}^{\text{qd}}(d{=}3) &= 4\times27+7\times50 = 458, \\
Q_{\text{register}}^{\text{qb}}(d{=}5\text{ instance}) &= 19\times50 = 950, \\
Q_{\text{register}}^{\text{qd}}(d{=}5) &= 4\times27+7\times50 = 458.
\end{align}
The notations $Q_{\text{register}}^{\text{qb}}$,$Q_{\text{register}}^{\text{qd}}$ refers to the physical qubit or qudit count for the surface code (excluding MSD) denoted by the subscript \textit{register}.
The hybrid implementations require fewer physical carriers (physical qubit or qudits) than their qubit-encoded counterparts. These numbers are further enhanced while counting the resources for the MSD next.

The efficiency of a MSD protocol is given by the yield parameter, $\gamma(d)$. This determines the scaling of the number of raw, noisy magic states with the acceptable error rate $\epsilon$. The number of raw copies of magic states required per output state scales as
\begin{equation}
N_{\text{raw}}(d) \propto \left[\ln(1/\epsilon)\right]^{\gamma(d)}.
\label{Nraw}
\end{equation}
A lower value of $\gamma$ corresponds to fewer resources needed to reach the same level of error. In the estimates below, we neglect the protocol-dependent proportionality constants in Eq.~\ref{Nraw}, since the purpose is to obtain an order-of-magnitude resource estimate rather than an exact cost. Qubits ($d=2$) use the 15-to-1 protocol with $\gamma(2) \approx 2.46$~\cite{bravyi2012magic}. Qutrits ($d=3$) use the $[[20,7,2]]_3$ protocol with $\gamma(3) = 1.51$~\cite{prakash2025low}, which is more efficient than all the known qubit triorthogonal codes of comparable size. For $d=5$, the $[519,106,5]_5$ code gives $\gamma(5) = 0.99$~\cite{saha2026sublogarithmic} that is a significant improvement as this belongs to the sublogarithmic-overhead regime. The asymptotic yield parameter for prime dimension $d$ approaches to $1/\ln d$ as $d\to\infty$~\cite{saha2026sublogarithmic}, so the overhead reduction is expected to continue with increasing $d$.
The total magic-state distillation cost $Q_{cost}^M$ for a given implementation, i.e. the total number of raw magic states consumed is given by,
\begin{equation}
Q_{cost}^M(d) = N_T(d) \times N_{\text{raw}}(d).
\end{equation}

For the example problem, we perform a MSD cost analysis now.
With an error rate $\epsilon=10^{-12}$, $\ln(1/\epsilon)\approx27.6$, using Eq.~\ref{Nraw} gives,
\begin{align}
N_{\text{raw}}(2) &\propto 27.6^{2.46} \approx 3520, \\
N_{\text{raw}}(3) &\propto 27.6^{1.51} \approx 150, \\
N_{\text{raw}}(5) &\propto 27.6^{0.99} \approx 27.
\end{align}

The corresponding MSD costs are
\begin{align}
Q_{cost}^M(d{=}2\text{, for }d{=}3\text{ instance}) &= 2014\times3520 \nonumber \\&\approx 7.09\times10^6, \\
Q_{cost}^M(d=3) &= 254\times150 \nonumber\\&\approx 3.81\times10^4, \\
Q_{cost}^M(d{=}2\text{, for }d{=}5\text{ instance}) &= 4945\times3520 \nonumber\\&\approx 1.74\times10^7, \\
Q_{cost}^M(d=5) &= 273\times27 \nonumber\\&\approx 7.37\times10^3.
\end{align}

In practice, MSD is not performed in isolation but through dedicated MSD factories. A MSD factory is a block of physical qubits/qudits running a distillation protocol continuously, so that instead of distilling one batch of magic states, it produces a multiple clean magic states that the fault-tolerant computation consumes as needed \cite{litinski2019game}.
The total physical resource estimate for an implementation at dimension $d$ is done by calculating the register cost using the surface code and MSD-factory costs,
\begin{equation}
Q_{\text{total}}(d) = Q_{\text{register}}(d) + Q^M_{factory}(d),
\end{equation}
where $Q_{\text{register}}(d)$ is built from $\kappa(\delta(d))$. $Q^M_{factory}(d)$ is the actual MSD cost when operated in a MSD factory. In general, it is performed in parallelized in two ways, number of factories working in parallel and the fraction of each factory that is active at a time. These factors help to produce high fidelity magic states faster and is an active area of research. For simplicity, let us assume $Q^M_{factory}(d) \propto Q_{cost}^M(d)$ which means a comparable factory parallelization efficiency across implementations for all $d$.

The total physical resoruces is thus,
\begin{align}
Q_{\text{total}}^{\text{qb}}(d{=}3\text{ instance}) &\approx 750+7.09\times10^6 \approx 7.09\times10^6, \nonumber\\
Q_{\text{total}}^{\text{qd}}(3) &\approx 458+3.81\times10^4 \approx 3.86\times10^4,\nonumber \\
\frac{Q_{\text{total}}^{\text{qb}}}{Q_{\text{total}}^{\text{qd}}}\bigg|_{d=3} &\approx 184, \\[4pt]
Q_{\text{total}}^{\text{qb}}(d{=}5\text{ instance}) &\approx 950+1.74\times10^7 \approx 1.74\times10^7, \nonumber\\
Q_{\text{total}}^{\text{qd}}(5) &\approx 458+7370 \approx 7828, \nonumber\\
\frac{Q_{\text{total}}^{\text{qb}}}{Q_{\text{total}}^{\text{qd}}}\bigg|_{d=5} &\approx 2223.
\end{align}
Even with this conservative estimate, the hybrid implementation offers a substantial reduction in total physical resources at both $d=3$ and $d=5$, with an order of magnitude increase in the advantage between the two. The resulting estimates depend on the specific MSD model considered here, which is chosen as a simplified model to show the order-of-magnitude differences in resource requirements.

As mentioned, the estimates assume same factory parallelization efficiency for all $d$ values i.e., the same number of parallel factories $F$ and the same total distillation time budget $T$. Under this assumption, the raw-state production rate of a factory is defined by following the factory-parallelization in~\cite{litinski2019game},
\begin{equation}
R(d) = \frac{Q_{cost}^M(d)}{F\,T}.
\end{equation}
Taking the $d=2$ (qubit) implementation as the reference, the relative rate required at dimension $d$ to match the qubit implementation's total output within the same time and factory budget is
\begin{equation}
\frac{R(d)}{R(2)} = \frac{Q_{cost}^M(d)}{Q_{cost}^M(2)}.
\end{equation}
Using the numbers above,
\begin{align}
\frac{R(3)}{R(2)} &\approx \frac{3.81\times10^4}{7.09\times10^6} \approx \frac{1}{186}, \\
\frac{R(5)}{R(2)} &\approx \frac{7.37\times10^3}{1.74\times10^7} \approx \frac{1}{2361},
\end{align}
under an equal-parallelization assumption, same $F$ for all $d$. This indicates that the qudit resource advantage does not depend on qudit distillation hardware being better or similar to the qubit factory's capabilities. A slower and less demanding qudit factory is enough to perform better than the qubit-based MSD. This motivates towards a more near-term-plausible hardware requirement for qudits to match qubit-factory output.

In summary, smaller code distance, fewer non-Clifford gates and more efficient MSD leads to the resource advantage for the native qudit-hybrid implementation over the all-qubit encoded algorithm for solving the IP problems. There is an overall resource reduction of over two orders of magnitude at $d=3$ and over three orders of magnitude at $d=5$ for the example problem, with the advantage continuing to grow for higher prime dimensions as $\gamma\to1/\ln d$ \cite{saha2026sublogarithmic}.

\section{Entanglement study for a hybrid qudit-qubit algorithm}
\label{Entanalg}

In this section, we characterize the quantum nature of the hybrid qudit-qubit algorithm by quantifying the entanglement and studying the entanglement structure. Unlike the fault-tolerant resource comparison in the previous section, the entanglement analysis is not restricted to prime dimensions, hence, $d=2,3,4$ are used in this section.
The two subsystems, qudit- and qubit- registers, provides a natural cut for studying the entanglement generation using the von-Neumann entropy as a measure. This also helps in understanding the entanglement generation in each part of the algorithm as the entanglement depends on the structure of the problem that is being solved.
The presence of global entanglement in the system makes it difficult to efficient classical simulate via tensor networks with bounded bond dimension \cite{schuch2008entropy}.
Hence, the bi-, tri- and four- partite mutual information is calculated as together they suggest global entanglement in the state.
The mutual information of higher order probes how entanglement is distributed among multiple subsystems simultaneously and it can certify the presence of genuine multipartite entanglement that cannot be decomposed into pairwise components \cite{cerf1998information,seshadri2018tripartite,vedral2002role,caceffo2023negative}. 
Furthermore, a numerical analysis to observe the growth of entanglement with the subsystem size (mixed qubits and qudits in one subsystem) is performed, which strongly hints towards a volume law-like behavior of the entanglement. 
The bipartite volume law is necessary for classical hardness, however, it is not sufficient. There are certain efficiently simulable states (e.g., stabilizer states) that exhibit volume-law entanglement \cite{gottesman1998heisenberg,sharma2025multipartite,nezami2020multipartite,aaronson2004improved,dowling2603classical}. The presence of multipartite entanglement (mutual information) and a volume-law behavior (von-Neumann entropy with subsystem size) together make a strong case for hardness in classical simulability of the hybrid quantum algorithm.  

\subsection{Entanglement generation: von Neumann Entropy across qudit-qubit cut}
As a measure of entanglement and to show that the algorithm generates structured entanglement, the von-Neumann entropy is studied in this section, defined as follows.
Consider a pure state $\ket{\psi}$ on the bipartite Hilbert space $\mathcal{H}_A \otimes \mathcal{H}_B$, where $\mathcal{H}_A$ may contain both qudits and qubits. The reduced density matrix of subsystem $A$ is
\begin{equation}
\rho_A = \Tr_B(\dyad{\psi}).
\end{equation}
The bipartite entanglement is quantified by the von Neumann entropy
\begin{equation}
S(A) = -\Tr(\rho_A \log \rho_A) = -\sum_i \lambda_i \log \lambda_i,
\label{eq:vn_entropy}
\end{equation}
where $\lambda_i$ are the eigenvalues of $\rho_A$. For a maximally entangled state across a cut with $\dim(\mathcal{H}_A) = D_A \leq D_B$, the entropy saturates at
\begin{equation}
S_{\max} = \log D_A.
\label{eq:max_entropy}
\end{equation}

\begin{figure*}[t]
\centerline{\includegraphics[width = 1\linewidth,trim={0cm 8cm 0cm 8cm},clip]{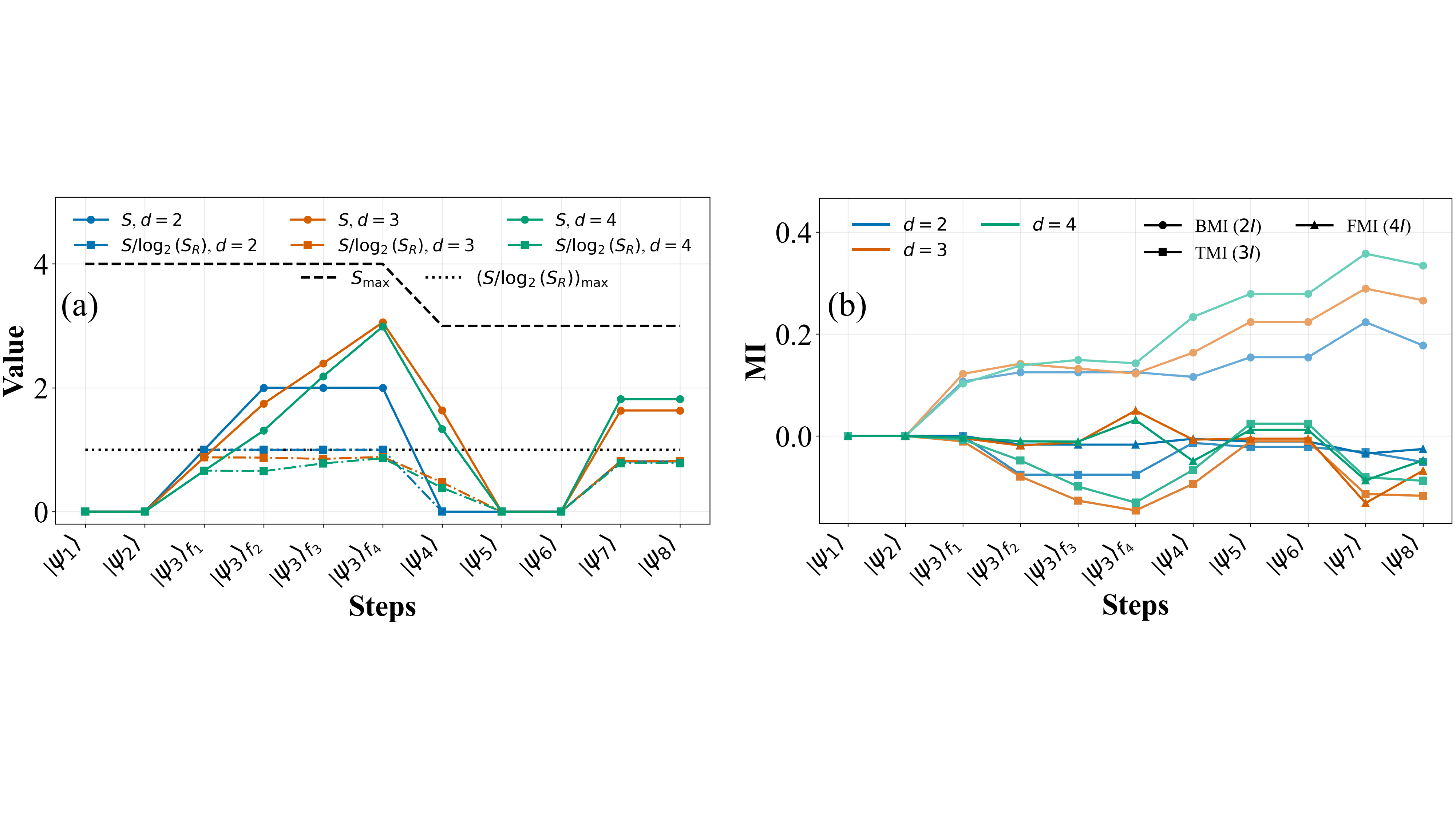}}
\caption{Entanglement study for solving the example problem given by Eq.~\eqref{ExP8} (a) von Neumann entropy $S$ across the qudit-qubit bipartition (solid lines) and the normalized ratio $S/\log_2(S_R)$ (dash-dotted lines) at each step of the algorithm, for qudit dimension $d=2$ (blue), $d=3$ (orange), and $d=4$ (green). The dashed black line marks the dimension-imposed maximum entropy $S_{\max}$, and the dotted black line marks $(S/\log_2(S_R))_{\max} = 1$, the maximum entropy achievable given the Schmidt rank $S_R$ realized at that step. $S/\log_2(S_R)\to 1$ indicates a state that is maximally entangled for its own Schmidt rank, not relative to the full Hilbert-space dimension. States $|\psi_3\rangle_{f_1}$-$|\psi_3\rangle_{f_4}$ label the intermediate qudit-qubit state after each of the $m=4$ sequential constraint unitaries $\hat U_{f_i}$ is applied. (b) Bipartite (BMI, circles), tripartite (TMI, squares), and four-partite (FMI, triangles) mutual information as a function of algorithm step, for qudit dimension $d=2$ (blue), $d=3$ (orange), and $d=4$ (green), averaged over all subsystem partitions of the corresponding order. There is a color gradient for the curves belonging to each of the $d$ values, which ranges from BMI (lighter) to FMI (darker).}
\label{f1}
\end{figure*}

At first we consider qudit and qubit as two separate subsystems, where $\mathcal{H}_A = d^n$, $\mathcal{H}_B = 2^m$ for stage I and $\mathcal{H}_A = d^n$, $\mathcal{H}_B = 2^{l+1}$ for stage II of the algorithm. 
There are multiple states during the evolution of the algorithm, for each state the von Neumann entropy $S$ is calculated (see Appendices~\ref{Entang_derivation}~and~\ref{entpsi7} for the derivations) as shown below.
\begin{itemize}
    \item $\ket{\psi_1}$: $S = 0$ (product state)
    \item $\ket{\psi_2}$: $S = 0$ (superposition product state)
    \item $\ket{\psi_3}$: $S = \ln d^n - \frac{1}{d^n} \sum_{\gamma=0}^{m-1} K_\gamma \binom{m}{\gamma} \ln\binom{m}{\gamma} - \frac{F_s}{d^n} \ln F_s$
    \item $\ket{\psi_4}$: $S = 0$ (superposition product state of qudits) if the amplitude amplification is perfect, otherwise non-zero.
    \item $\ket{\psi_5}$: $S = 0$ 
    \item $\ket{\psi_6}$: $S = 0$ 
    \item $\ket{\psi_7}$: $S = -\sum_{j=1}^M \frac{d_j}{F_s} \ln \frac{d_j}{F_s}$ 
    \item $\ket{\psi_8}$: $S = -\sum_{j=1}^M \frac{d_j}{F_s} \ln \frac{d_j}{F_s}$ 
    \item $\ket{\psi_9}$: $S \leq \log{2}$ (after measuring the single qubit)
\end{itemize}
The system initializes in a product state $\ket{\psi_1}$ of two registers, $n$-qudits and $m$-qubits. $n$-Hadamard gates are applied to the qudit register to give $\ket{\psi_2}$. There is no entanglement ($S= 0$) between the qudit and the qubit registers in $\ket{\psi_1}$ and $\ket{\psi_2}$. In the next step, multiple unitary operators are applied to satisfy the constraints, which by construction acts on the qudit and the qubit register introducing entanglement between them leading to the state $\ket{\psi_3}$. $K_{\gamma}$ is the multiplicity of the state that is entangled with the qubit state consisting of $\gamma$ ones in the multi-qubit state. $\ket{\psi_4}$ is obtained after the amplitude amplification step and the measurement of the qubit register, $\ket{\psi_5}$ is the state where $l+1$ new qubits are introduced and $\ket{\psi_6}$ is obtained after applying $l$-Hadamard gates to the qubit register, all of which has zero entanglement between qudit-qubit cut. The algorithm correlates the qudit and the new qubit register again in the next step by imprinting qudit dependent phases onto the $l$-qubits in state $\ket{\psi_7}$, generating entanglement between the two registers. The same entanglement content is retained after the inverse QFT operation in $\ket{\psi_8}$. The single qubit is entangled with the rest of the system via a controlled rotation, given by $\ket{\psi_9}$, where the entanglement is bounded by $\log2$.  

The states that are most relevant from the entanglement perspective are $\ket{\psi_{3}}$ and $\ket{\psi_{7}}$.
The state $\ket{\psi_{3}}$ is obtained after the applications of $m$ constraint unitaries $U_{f_i}$, that separates the feasible and the infeasible region. This step determines the spatial entanglement between the qudits. The entanglement in $\ket{\psi_7}$ is used to extract the cost function of the feasible region that is maximized in the subsequent steps.  
The quantum phase estimation (QPE) routine introduces controlled applications of the cost operator $\hat{O}_c$, which is diagonal in the computational basis:
\begin{equation}
\hat{O}_c \ket{\mathbf{x}} = e^{2\pi i C(\mathbf{x})/C_{ub}} \ket{\mathbf{x}}.
\label{eq:cost_operator}
\end{equation}
$\hat{O}_c$ acts only on the qudit register and does not directly entangle qudits with qubits. However, the Hamiltonian simulation using $\hat{O}_c$ acts on both the registers, creates entanglement between the QPE register and the qudits. The QPE routine entangles $l$ qubits with the qudit register.    

During the analysis, we evaluate the Schmidt rank $S_R$ for each of the steps in the algorithm. The Schmidt rank $S_R = |\{k : \lambda_k > 0\}|$ counts the number of non-zero eigenvalues which is the exact number of orthogonal states in the qudit register that are correlated with distinct states of the qubit register. $S_R=1$ indicates a separable state, while a larger Schmidt rank indicates that more dimensions of the two
subsystems participate in the entanglement. For each step, the ratio $S/\log_2{S_R}$ of the von Neumann entropy $S$ and the Schmidt rank $\log_2{S_R}$  provides a measure of the average entanglement entropy per Schmidt component. While $\log_2{S_R}$ quantifies the number of dimensions participating in the entanglement, $S$ quantifies the overall
entanglement. Therefore, $S/\log_2{S_R}$ gives an indication of how uniformly the entanglement is distributed among the nonzero Schmidt coefficients. A larger value indicates a more even distribution of entanglement across the Schmidt components. The ratio is bounded $0 \leq S/\log_2{S_R}\leq 1$.

\noindent \textbf{Numerical results}:
For the numerical analysis, we look at the entanglement structure for different values $d$ of the integer variables $x_i$ across the algorithm. 
Fig.~\ref{f1} shows the von Neumann entropy across the qudit–qubit bipartition as the algorithm evolve across multiple steps. The problem described by Eq.~\ref{ExP8} (panel a) is solved for varying qudit dimensions $d=2,3,4$ (integer values), with $m=4$, $n=4$, $l=3$. The example problem is quartic, this plays a crucial role in determining the entanglement structure. The dimension of the subsystems impose upper bound (max entropy as shown in dashed black line), which is $\min\{2^m,d^n\}$ for stage I and $\min\{2^l,d^n\}$ for stage II.  In all the cases, the entropy grows steadily with the constraint unitary operators, reaching a maximum (below the maximum possible entanglement) before the Grover-amplification step, indicating structured entanglement between the two registers. The von-Neumann entropy reaches a maxima for $\ket{\psi_3}_{f_4}$ and for the state $\ket{\psi_4}$ it dips to near zero as expected from the analytical expression above. Finally, it goes to zero when the measurement of the $m$-qubit constraint register is performed, which collapses the qudit array onto the feasible subspace and disentangles it from the measurement outcome. The entropy further rises in the cost-optimization stage as the $l$-qubit register entangles with the feasible qudit state via quantum phase estimation. Similar qualitative behavior of the entanglement generation in the algorithm is observed for all the problems that were solved numerically, which are not shown in this work. Interestingly, the peak of entanglement in either of the stages can shift depending on the problem structure as shown in Appendix~\ref{entanpeak}. Larger entanglement is generated in the first stage of the algorithm when the problem is highly constrained while it is larger for the second stage for relatively less constrained problem (See Figs.~\ref{entfig1}~and~\ref{entfig2} in Appendix~\ref{entanpeak}).
As the dimension increases, the problem becomes more constrained, leading to stage I generating more entanglement with $d$. 
The ratio $S/\log_2{S_R}$ is shown to stay close to maximal value (dotted black line) in both the stages for all the cases, indicating that the entanglement is distributed among the Schmidt components. The maximal value is $1$ (dotted black line), which indicates that the system is maximally entangled, occupying all the available Schmidt components. The entanglement spectrum of the reduced density matrix of the qudit subsystem to probe the structure of the entanglement and track how scrambled the information is in the circuit is shown in Appendix~\ref{spectrumm}.

\subsection{Classical simulability: mutual information and global entanglement}
In order to study the global entanglement in the system to give evidence of the difficulty in classical simulability of the algorithm, mutual information of various orders is studied here.
For a partition of the system into $K$ disjoint subsystems $A_1, \dots, A_K$, the $K$-partite mutual information is defined as $I_K = \sum_{j=1}^{K} (-1)^{j+1} \sum_{1 \leq i_1 < \cdots < i_j \leq K} S(A_{i_1} \cup \cdots \cup A_{i_j})$. For $K=2$, this reduces to the standard bipartite mutual information $I_2(A:B) = S(A) + S(B) - S(AB)$ (where $S$ is the von-Neumann entropy), which quantifies the total correlation between two subsystems. However, $I_2(A:B)$ does not distinguish between genuine multipartite entanglement and pairwise correlations. To go beyond bipartite correlations and establish global entanglement, multipartite mutual information (tripartite $I_3(A:B:C)$ and four-partite $I_4(A:B:C:D)$) are calculated. These higher-order mutual-information quantities lead to the presence and structure of multipartite correlations.

For a tripartite system $A:B:C$, the mutual information is
\begin{equation}
\begin{split}
I_3(A:B:C) &= S(A) + S(B) + S(C) \\ 
&- S(AB) - S(AC) - S(BC) + S(ABC).
\end{split}
\label{eq:tripartite_mi}
\end{equation}
A negative tripartite mutual information $I_3$ indicates a synergistic multipartite correlations. Whereas a positive $I_3$ leads to redundancy which can be expressed with only bipartite correlations \cite{bolter2023tripartite,caceffo2023negative}. 
Similarly, for a four-partite system $A:B:C:D$, the generalization is
\begin{equation}
I_4 = \sum_{i} S_i - \sum_{i<j} S_{ij} + \sum_{i<j<k} S_{ijk} - S_{ABCD}.
\label{eq:fourpartite_mi}
\end{equation}
A non-zero $I_4(A:B:C:D)$ leads to higher-order four-partite correlations. 
The multipartite mutual information $I_3$ and $I_4$ provide stronger witnesses, indicating entanglement that cannot be captured by pairwise correlations.

\noindent \textbf{Numerical results}: Fig.~\ref{f1}(b) show the bipartite (BMI), tripartite (TMI), and four-partite (FMI) mutual information across circuit depth averaged over all the combinations for solving the example problem given by Eq.~\ref{ExP8}. BMI follows the same growth–collapse–regrowth pattern as the von-Neumann entropy in Fig.\ref{f1}(a), which also captures the constraint-measurement step at which the qubit and qudit registers are disentangled. The TMI becomes negative during Stage I, indicating the build up of genuinely synergistic, multipartite correlations, rather than redundant pairwise correlations for all the values of $d$. These tripartite synergistic correlations dominate the entanglement structure are generated by the constraint satisfaction step, however, FMI is non-zero in the system. Higher-than-tripartite correlations contribute less to the global entanglement structure for problems with linear, qudratic and cubic terms. Given the quartic structure of the example problem FMI also contribute significantly. Taken together, the BMI, TMI and FMI, the algorithm generates predominantly bipartite, tripartite-synergistic and four-partite correlations in the system. Higher-order MI can be studied to further characterize the correlations in the algorithm for specific problems, which is out of the scope for this work.

The hierarchy $|I_2| > |I_3| > |I_4| > 0$ observed across all the cases indicates the entanglement structure generated by the hybrid algorithm for the particular problem at hand. The dominant correlations are bipartite, but there exists higher-order correlations that cannot be truncated or ignored. This is non-trivial as even if each of the constraint functions $C_i$ in the problem depend on two of the four variables, however, together they could create correlations that couple all variables simultaneously rather than in isolated pairs. The negative $I_3$ in our algorithm, places it outside the class of efficiently simulable Clifford circuits as the non-Clifford gate content established in Sec.~\ref{cliffnocliff} independently rules out stabilizer-based simulation. The additional non-zero $I_4$ further strengthens this conclusion. 
Hence, it is difficult to represent these states as a low-order tensor network with constant bond dimension, even with architectures designed to capture multipartite entanglement. 

\subsection{Growth of entanglement: von Neumann entropy with varying subsystem size}

To characterize the classical simulability of the hybrid qudit-qubit algorithm, one of the necessary conditions is to understand how the entanglement (von Neumann entropy) scales with the subsystem for which we perform a numerical analysis. A subsystem is defined by a mix of qudits and qubits, where the size is determined by the dimension of the Hilbert space spanned by the qudits and qubits. 
The system for the first stage consists of $n$-qudits and $m$-qubits, whereas any subsystem consists of $n_A$ qudits and $m_A$ qubits with the corresponding subsystem size given by $D_A = d^{n_A} \cdot 2^{m_A}$.
For a given subsystem size, there exists multiple different combinations of qudits and qubits, all of which are considered for calculating the von-Neumann entropy as a function of $D_A$ and averaged over all of them.
A similar analysis is also done for Stage~II of the algorithm with $n$-qudits and $l$-qubits.

As expected, the entanglement in the algorithm is generated from the application of the $m$ constraint unitaries in Stage~I. Each unitary $U_{f_i}$ implements a controlled permutation leading to a global dependence of multiple qudits and qubits after all $m$ constraints are applied. Any qubit or qudit that remains in the subsystem after tracing out the rest retains correlations through the constraint functions. With the subsystem dimension $D_A$, more of these correlations become visible as entanglement entropy.
In Stage~II, the cost operator $\hat{O}_c$ is diagonal in the computational basis and acts only on the qudit register. Hence, it imprints a phase but does not change the entanglement structure. However, when controlled versions of $\hat{O}_c$ are used for Hamiltonian simulation in the QPE routine, it entangle the $l$ QPE qubits with the qudit register. For the subsystem scaling analysis, the focus is on the state after Stage~I plus QPE which contains the complete entanglement structure of the algorithm.

\begin{figure*}[t]
\centerline{\includegraphics[width = 1\linewidth,trim={0cm 6.5cm 0cm 7cm},clip]{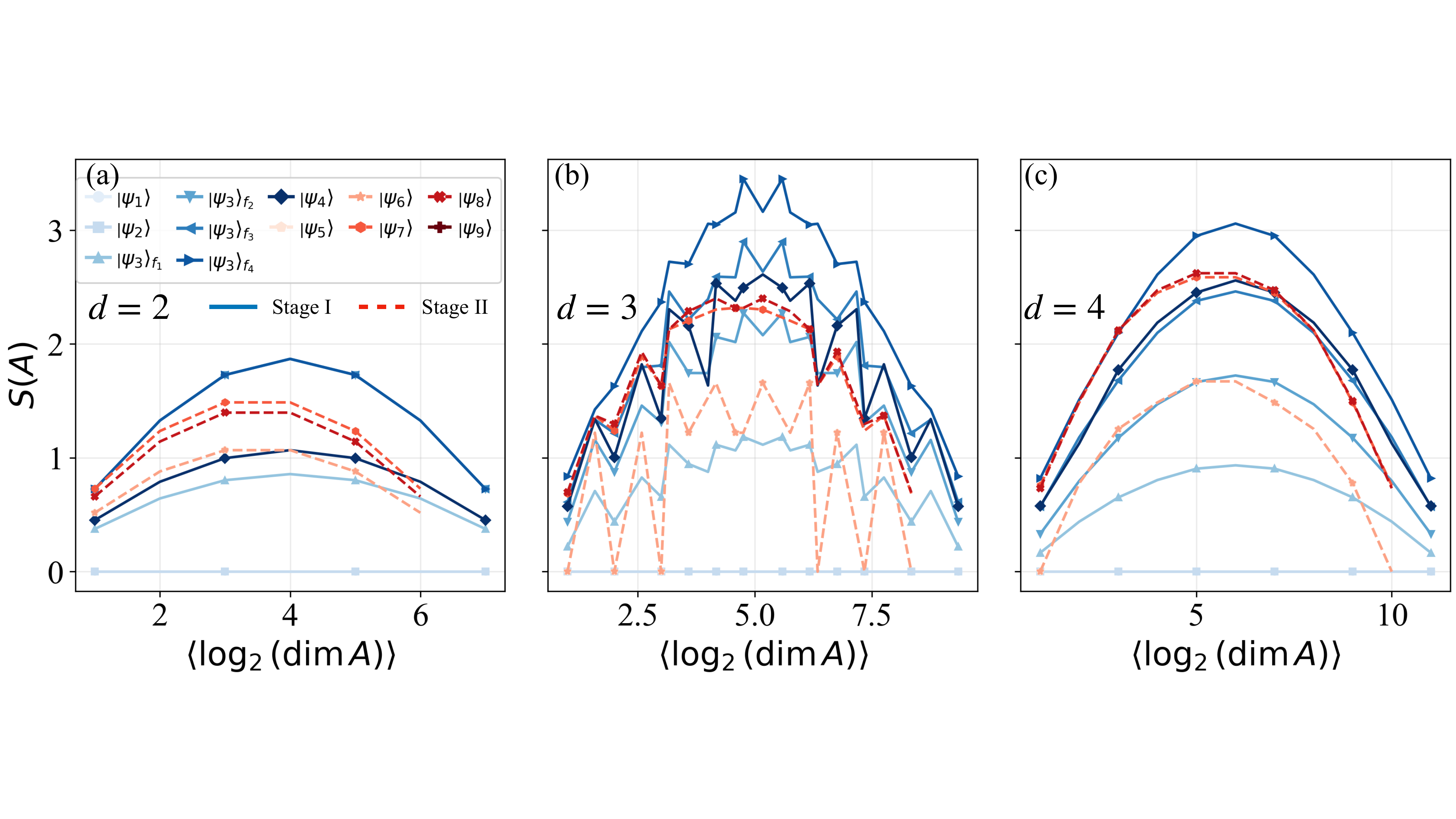}}
    \caption{Growth of the reduced-state entropy $S(A)$ with subsystem size $\langle\log_2(\dim A)\rangle$, averaged over all subsystems of a given size formed from qudits and qubits, for $d=2$ [(a)], $d=3$ [(b)], and $d=4$ [(c)]. Solid curves are Stage-I states and dashed curves are Stage-II states. The growth curves are calculated while solving the example problem given by Eq.~\eqref{ExP8}.}
\label{f2}
\end{figure*}

\noindent \textbf{Numerical results}: Fig.~\ref{f2} shows the growth of the entropy $S(A)$ with subsystem size $\log_2(dim A)$, for subsystems consisting of qudits and qubits. The initial state is a product state, which trivially leads to vanishing entropy across all the cuts. The middle panel (b) shows the growth of entanglement for a hybrid qutrit-qubit system, and the odd dimension ($d=3$) qudit leads to the staggered growth. This growth pattern is emerges when each unit (qutrit or qubit) added to the subsystem, it stores the information non-uniformly ($\log{3}$ or $\log{2}$). As the entanglement is generated in the system, the subsequent states exhibit a Page-like curve \cite{PhysRevLett.71.1291} where the entropy increases with subsystem size, peaks at the half-system cut, and decreases symmetrically. The peaks are higher for $d=3$ (panel b) and $d=4$ (panel c) as compared to the $d=2$ (panel a) case for the example problem. This shows that when the Hilbert space grows with $d$, the amount of information stored in the system also grows, indicating that the algorithm entangles larger parts of the available Hilbert space, rather than saturating.   
This scaling in all the plots is a signature of the volume law, where the entanglement entropy is proportional to the logarithm of the subsystem dimension (equivalent to the number of qudits or qubits in the subsystem), which is the number of degrees of freedom in the subsystem. In terms of the number of qudits and qubits,
\begin{equation}
\langle S(n_A, m_A) \rangle \propto n_A \log d + m_A \log 2,
\label{eq:vn_linear_hybrid}
\end{equation}
indicating that each qudit contributes $\propto\log d$ and each qubit contributes $\propto\log 2$ to the total entanglement, up to the saturation limit set by the dimension of the subsystems.
For subsystems where $D_A$ approaches $D_{\text{total}}/2$, the entropy saturates at the Page value \cite{PhysRevLett.71.1291}:
\begin{equation}
S_{\text{sat}} \approx \log D_A - \frac{D_A}{2 D_{\bar{A}}},
\label{eq:page_hybrid}
\end{equation}
and then decreases symmetrically for $D_A > D_{\text{total}}/2$ due to the global state being pure. A similar behavior is observed for the stage II of the algorithm. 
The flatness of the growth would indicate a area-law behavior which is not the case here.

This volume law behavior has direct consequences for classical tensor network simulation. For a Matrix Product State (MPS) representation of the hybrid system, the bond dimension $\chi$ across any bipartition must satisfy $\chi \gtrsim e^{\langle S \rangle}$.
With the volume law $\langle S \rangle \propto \log D_A$, the bond dimension at a cut where $D_A \approx D_{\text{total}}/2$ scales as:
\begin{equation}
\chi \gtrsim e^{c \log D_{\text{total}}} \propto D_{\text{total}}^{c},
\label{eq:chi_power}
\end{equation}
where $c \approx 1/2$ for the half-cut. For $n_A$ qudits of dimension $d$ and $m_A$ constraint qubits, this gives,
\begin{equation}
\chi \gtrsim d^{n_A/2} \cdot 2^{m_A/2},
\label{eq:chi_explicit}
\end{equation}
which is exponential in the number of variables $n_A$ and $m_A$.
The classical MPS simulation resources therefore scale as:
\begin{align}
\text{Memory} &\propto \chi^2 \gtrsim d^{n_A} \cdot 2^{m_A}, \label{eq:memory_exp}\\
\text{Time} &\propto \chi^3 \gtrsim d^{3n_A/2} \cdot 2^{3m_A/2}. \label{eq:time_exp}
\end{align}
Both are exponential in the system size. For $n = 20$, $d = 5$, and $m = 10$ constraint, the minimum bond dimension is $\chi \gtrsim 5^{10} \cdot 2^5 \approx 3.1 \times 10^8$, requiring exabytes of memory. For $n = 30$, the requirements exceed any foreseeable classical computing capacity.
Extrapolating the observed volume-law-like entanglement scaling to larger \(n\) suggests an exponentially increasing cost for MPS-based classical simulation, providing further evidence for increasing classical simulation difficulty.

\section{Discussion and Conclusions}
\label{discuss}
Two main questions concerning the hybrid qudit-qubit algorithm for integer programming are addressed in this work. The first one is regarding the resource comparison of the hybrid qudit-qubit algorithm with its qubit-only implementation that justifies the usage of qudits for the problem, and the second one is about the classical simulability of the algorithm by studying the entanglement generated in the system. We take an exemplar non-linear IP problem (quartic) that is solved using the quantum algorithm.

The resource analysis is performed to discuss the potential advantages of the hybrid algorithm over the qubit based encoding. This is done at the logical gate level by comparing the non-Clifford gate counts and then at the physical level by comparing the resources needed for a fault tolerant implementation of the algorithm in both the cases (qudit hybrid vs qubit only). At the logical gate level, up to leading order estimates, the qutrit-hybrid implementation required $7$-$9\times$ fewer total gates and non-Clifford gates than the qubit encoding case for the example problem. This resource efficiency increases further in the fault tolerance regime. The qutrit magic-state distillation protocol has a lower yield parameter ($\gamma = 1.51$\cite{prakash2025low} versus $\gamma = 2.46$\cite{bravyi2012magic} for the qubit protocols) which reduces the number of raw magic states required per logical non-Clifford gate by a factor of $\sim\!23$ at a target error rate of $\epsilon = 10^{-12}$. The total resource estimate for the example problems are two orders of magnitude lesser as compared to the qubit case within the simplified resource counting model. This raw magic state resource advantage over all-qubit encoding grows to $\sim\!130$ times when compared with a ququint ($d=5$ with $\gamma =0.99$ \cite{saha2026sublogarithmic}) implementation of the algorithm.

The study of the source of advantage led us to various parameters to discuss the entanglement generated in the algorithm. It is established that the algorithm generates entanglement across the qudit-qubit cut, with the dominant contribution from the constraint block and the QPE block. The QPE block contributes more when the problem is less constrained leading to finding the optimal solution from a larger feasible subspace, which is observed in the generation of more entanglement in stage II. The ratio of the von-Neumann entropy to the log of Schmidt rank remains close to maximal throughout the algorithmic run which indicates that the entanglement is distributed close to uniformly among the Schmidt components that are populated.
The growth of the entanglement with system size shows a Page-like volume law behavior, which is a necessary but not sufficient condition for classical hardness of the intermediate states. The evidence for a volume-law behavior is further strengthened by the multipartite mutual information that indicates global entanglement in the system. The negative tripartite mutual information generated during the constraint-satisfaction step, shows that this entanglement is synergistic in nature.

These two results are crucial for near-term qudit hardware development. The observed entanglement structure provides evidence of increasing difficulty for classical tensor-network simulations with bounded bond dimension and comparatively cheaper to run with qudit hardware. Together with the non-Clifford content, the non-zero higher-order correlations provide additional evidence that simple stabilizer- or low-order correlation-based simulation may be insufficient. It is the combination of the two results that makes a case for pursuing native qudit control as a fault-tolerant computing alternative, at least for optimization problems. 
On the hardware side, the resource count of this work is relevant if it can be demonstrated on a qudit platform. Experimentally, there are several candidates that can verify the resource advatange, including trapped-ion qudits~\cite{ringbauer2022universal,shi2026efficient}, neutral-atom and Rydberg-dressed qudit arrays~\cite{burshtein2026robust}, and photonic qudit encodings~\cite{Chi2022,Kim2024}. Extending the entanglement and mutual-information analysis to larger $n$ numerically can provide stronger arguments for the classical-hardness of simulating the algorithm.

\begin{acknowledgments}
This work is funded by the German Federal Ministry of Education and Research within the funding program “Quantum Technologies - from basic research to market” under Contract No. 13N16138.
\end{acknowledgments}

\section*{Data Availability}

All data supporting the findings of this study are included in the article.

\onecolumngrid
\newpage
\appendix

\setcounter{equation}{0}
\setcounter{figure}{0}
\setcounter{table}{0}
\setcounter{section}{0}
\renewcommand{\thesection}{A\arabic{section}}
\renewcommand{\theequation}{A\arabic{equation}}
\renewcommand{\thefigure}{A\arabic{figure}}

\section{Gate Count Calculation: Example problem}
\label{gatecount}
We demonstrate the gate count scaling using the integer programming problem from Eq.~(\ref{ExP8}), using the circuit parameters:
\begin{equation}
n = 4, \quad m = 4, \quad d = 3, \quad k = \lceil \log_2 d \rceil = 2, \quad l = 3.
\end{equation}
The cost function has $M = 14$ monomials with average body-order $\bar{r} = 29/14 \approx 2.071$. The base constants are:
\begin{equation}
d^{n/2} = 9, \quad \log_2 d \approx 1.585, \quad 2^l = 8, \quad l^2 = 9.
\end{equation}
The total gate count for the algorithm is:
\begin{equation}
G_{\text{total}} = G_{U_f} + G_{\text{amp}} + 2^l \cdot G_{O_c} + G_{\text{QFT}},
\end{equation}
where the cost operator is applied $2^l$ times total: once during state preparation and $2^l-1$ times during QPE.

\subsection{Constraint Unitaries $\hat{U}_f$}
All-to-all:
\begin{align}
G_{U_f}^{\text{qd}} &= m \cdot n = 16, \\
G_{U_f}^{\text{qb}} &= m \cdot n(\log_2 d)^2 \approx 40.
\end{align}
Nearest-neighbor:
\begin{align}
G_{U_f}^{\text{qd}} &= m \cdot n^2 = 64, \\
G_{U_f}^{\text{qb}} &= m \cdot n^2(\log_2 d)^4 \approx 404.
\end{align}

\subsection{Amplitude Amplification}
All-to-all:
\begin{align}
G_{\text{amp}}^{\text{qd}} &= d^{n/2} \cdot n = 36, \\
G_{\text{amp}}^{\text{qb}} &= d^{n/2} \cdot n(\log_2 d)^3 \approx 143.
\end{align}
Nearest-neighbor:
\begin{align}
G_{\text{amp}}^{\text{qd}} &= d^{n/2} \cdot n^2 = 144, \\
G_{\text{amp}}^{\text{qb}} &= d^{n/2} \cdot n^2(\log_2 d)^4 \approx 908.
\end{align}

\subsection{Cost Operator $\hat{O}_c$ (State Preparation)}
All-to-all:
\begin{align}
G_{O_c}^{\text{qd}} &= M \cdot \bar{r} = 14 \times 2.071 \approx 29, \\
G_{O_c}^{\text{qb}} &= M \cdot \bar{r}^2 k^2 = 14 \times (2.071)^2 \times 4 \approx 240.
\end{align}
Nearest-neighbor:
\begin{align}
G_{O_c}^{\text{qd}} &= M \cdot \bar{r} \cdot n = 29 \times 4 = 116, \\
G_{O_c}^{\text{qb}} &= M \cdot \bar{r}^2 k^2 \cdot n \approx 240 \times 4 \approx 959.
\end{align}

\subsection{QPE (Controlled-$\hat{O}_c$ + QFT)}
All-to-all:
\begin{align}
G_{\text{QPE}}^{\text{qd}} &= 7 \times 29 + 9 = 212, \\
G_{\text{QPE}}^{\text{qb}} &= 7 \times 240 + 9 = 1689.
\end{align}
Nearest-neighbor:
\begin{align}
G_{\text{QPE}}^{\text{qd}} &= 7 \times 116 + 9 = 821, \\
G_{\text{QPE}}^{\text{qb}} &= 7 \times 959 + 9 = 6722.
\end{align}

\subsection{Total Gate Counts}
All-to-all:
\begin{align}
G_{\text{total}}^{\text{qd}} &= 16 + 36 + 8 \times 29 + 9 = 293, \\
G_{\text{total}}^{\text{qb}} &= 40 + 143 + 8 \times 240 + 9 = 2112, \\
\frac{G_{\text{total}}^{\text{qb}}}{G_{\text{total}}^{\text{qd}}} &= \frac{2112}{293} \approx 7.21.
\end{align}
Nearest-neighbor:
\begin{align}
G_{\text{total}}^{\text{qd}} &= 64 + 144 + 8 \times 116 + 9 = 1145, \\
G_{\text{total}}^{\text{qb}} &= 404 + 908 + 8 \times 959 + 9 = 8993, \\
\frac{G_{\text{total}}^{\text{qb}}}{G_{\text{total}}^{\text{qd}}} &= \frac{8993}{1145} \approx 7.85.
\end{align}

\subsection{Non-Clifford Gate Counts}
All-to-all:
\begin{align}
N_T^{\text{qd}} &= 0.33(16) + 0.29(36) + 1.00(29) + 0.99(212) = 254, \\
N_T^{\text{qb}} &= 0.47(40) + 0.46(143) + 1.00(240) + 1.00(1689) = 2014, \\
\frac{N_T^{\text{qb}}}{N_T^{\text{qd}}} &= \frac{2014}{254} \approx 7.93.
\end{align}
Nearest-neighbor:
\begin{align}
N_T^{\text{qd}} &= 0.33(64) + 0.29(144) + 1.00(116) + 0.99(821) = 992, \\
N_T^{\text{qb}} &= 0.47(404) + 0.46(908) + 1.00(959) + 1.00(6722) = 8289, \\
\frac{N_T^{\text{qb}}}{N_T^{\text{qd}}} &= \frac{8289}{992} \approx 8.36.
\end{align}

\section{Clifford vs Non-Clifford Gate Classification}
\label{cliffnocliff}
\begin{center}
\begin{tabular}{@{}lcc@{}}
\toprule
\textbf{Gate} & \textbf{Qudit (Clifford/Non-Clifford)} & \textbf{Qubit (Clifford/Non-Clifford)} \\
\midrule
SUM / CNOT & Clifford & Clifford \\
SWAP (3$\times$ SUM/CNOT) & Clifford & Clifford \\
Hadamard & Clifford & Clifford \\
$R_z(\theta)$ arbitrary & --- & Non-Clifford \\
$R(\theta,\phi)$ arbitrary & Non-Clifford & --- \\
Controlled-phase $C\!P(\theta)$ & Non-Clifford & Non-Clifford \\
$T$ gate & --- & Non-Clifford \\
Toffoli (decomposed) & --- & 6 CNOT (Clifford) + 7 $T$ (Non-Clifford) \\
Generalized Toffoli & $\mathcal O(1)$ Non-Clifford per transposition & $\mathcal O(nk)$ Non-Clifford gates \\
\bottomrule
\end{tabular}
\end{center}

\section{Entanglement for $\ket{\psi_3}$ derivation}
\label{Entang_derivation}

\noindent The full state is
\begin{equation}
\ket{\Psi} = \frac{1}{d^{n/2}} \sum_{\gamma=0}^{m-1} \sqrt{\binom{m}{\gamma}} \sum_{k=1}^{K_\gamma} \ket{y_k^{(\gamma)}}_{\text{qd}} \otimes \ket{e^\gamma_k}_{\text{qb}} 
+ \frac{1}{d^{n/2}} \sum_{i=1}^{F_s} \ket{\psi_i}_{\text{qd}} \otimes \ket{11\cdots 1}_{\text{qb}}
\end{equation}
with orthonormality conditions
\begin{align}
\braket{y_k^{(\gamma)}}{y_{k'}^{(\gamma')}} &= \delta_{\gamma\gamma'}\delta_{kk'}, \quad
\braket{\psi_i}{\psi_j} = \delta_{ij}, \quad
\braket{e^\gamma_k}{e^{\gamma'}_{k'}} = \delta_{\gamma\gamma'}\delta_{kk'} \\
\braket{11\cdots 1}{e^\gamma_k} &= 0, \quad
\braket{\psi_i}{y_k^{(\gamma)}} = 0. \nonumber
\end{align}
Tracing out the qubits gives the reduced density matrix
\begin{equation}
\rho_{\text{qd}} = \Tr_{\text{qb}}(\ket{\Psi}\bra{\Psi}) 
= \frac{1}{d^n} \sum_{\gamma=0}^{m-1} \binom{m}{\gamma} \sum_{k=1}^{K_\gamma} \ket{y_k^{(\gamma)}}\bra{y_k^{(\gamma)}}
+ \frac{1}{d^n} \sum_{i,j=1}^{F_s} \ket{\psi_i}\bra{\psi_j}.
\end{equation}
The second term simplifies to
\begin{equation}
\sum_{i,j=1}^S \ket{\psi_i}\bra{\psi_j} = \ket{\Psi_m}\bra{\Psi_m}, \quad \ket{\Psi_m} = \sum_{i=1}^S \ket{\psi_i}, \quad \braket{\Psi_m}{\Psi_m} = {F_s}.
\end{equation}
The normalized eigenvector is \(\ket{\phi_m} = \ket{\Psi_m}/\sqrt{{F_s}}\), giving eigenvalue \({F_s}/d^n\) with multiplicity 1.
The first term is already diagonal in the \(\{\ket{y_k^{(\gamma)}}\}\) basis:
\begin{equation}
\lambda_{\gamma,k} = \frac{\binom{m}{\gamma}}{d^n}, \quad \text{multiplicity } K_\gamma \text{ for each } \gamma.
\end{equation}
The full eigenvalue spectrum is
\begin{equation}
\left\{ \underbrace{\frac{\binom{m}{\gamma}}{d^n} \times K_\gamma}_{\gamma=0,\ldots,m-1}, \;\; \frac{{F_s}}{d^n} \times 1, \;\; 0 \times \left(d^n - \sum_{\gamma=0}^{m-1} K_\gamma - 1\right) \right\}.
\end{equation}
Normalization \(\Tr(\rho_{\text{qd}}) = 1\) requires
\begin{equation}
\sum_{\gamma=0}^{m-1} K_\gamma \binom{m}{\gamma} + {F_s} = d^n.
\end{equation}
The von Neumann entropy is
\begin{align}
S(\rho_{
\text{qd}}) &= -\sum_{\gamma=0}^{m-1} K_\gamma \frac{\binom{m}{\gamma}}{d^n} \ln\left( \frac{\binom{m}{\gamma}}{d^n} \right) - \frac{{F_s}}{d^n} \ln\left( \frac{{F_s}}{d^n} \right) \\
&= \ln d^n - \frac{1}{d^n} \sum_{\gamma=0}^{m-1} K_\gamma \binom{m}{\gamma} \ln\binom{m}{\gamma} - \frac{F_s}{d^n} \ln {F_s}.
\end{align}

\subsection*{Limiting Cases}

\textbf{Case 1: No feasible space (\({F_s} = 0\)).}
\begin{equation}
\sum_{\gamma=0}^{m-1} K_\gamma \binom{m}{\gamma} = d^n, \quad S = \ln d^n - \frac{1}{d^n} \sum_{\gamma=0}^{m-1} K_\gamma \binom{m}{\gamma} \ln\binom{m}{\gamma}.
\end{equation}
If all weight is at \(\gamma = 0\) or \(\gamma = m-1\) where \(\binom{m}{\gamma}=1\):
\begin{equation}
{F_s} = \ln d^n \quad \text{(maximal)}.
\end{equation}
If all weight is at \(\gamma = \lfloor m/2 \rfloor\):
\begin{equation}
S \approx \ln d^n - m\ln 2 + O(\ln m).
\end{equation}

\textbf{Case 2: Only feasible space (\(K_\gamma = 0\) for all \(\gamma\)).}
\begin{equation}
{F_s} = d^n, \quad \rho_{\text{qd}} = \frac{1}{d^n}\ket{\Psi_m}\bra{\Psi_m}, \quad S(\rho_{\text{qd}}) = \ln d^n - \frac{d^n}{d^n}\ln d^n = 0 \quad \text{(pure state)}.
\end{equation}

\textbf{Case 3: Equal distribution (\(K_\gamma = K\) constant).}
\begin{equation}
K = \frac{d^n - {F_s}}{2^m - 1}, \quad S \approx \ln d^n - m\ln 2 + \ln 2 - \frac{{F_s}}{d^n}\ln {F_s}.
\end{equation}

\textbf{Case 4: Large qudit dimension (\(d^n \gg 2^m\)).}
\begin{equation}
\frac{\binom{m}{\gamma}}{d^n} \to 0, \quad \frac{{F_s}}{d^n} \to 0, \quad S \to \ln d^n \quad \text{(approaches maximal entropy)}.
\end{equation}

\textbf{Case 5: Small qudit dimension (\(d^n \ll 2^m\)).}
\begin{equation}
S \leq \ln d^n \quad \text{(bounded by the qudit Hilbert space)}.
\end{equation}

\textbf{Bounds}
\begin{equation}
0 \leq S(\rho_{\text{qd}}) \leq \ln d^n
\end{equation}
\begin{equation}
S(\rho_{\text{qd}}) \leq \min\left(\ln d^n, \; m\ln 2\right) \quad \text{for typical distributions}.
\end{equation}

\section{Entanglement for $\ket{\psi_7}$ derivation}
\label{entpsi7}
\noindent The state after QPE is
\begin{equation}
\ket{\psi_7} = \tilde{F}_n \sum_{\mathbf{y}_s} \sum_{k=0}^{2^l-1} e^{i2\pi \tilde{\phi}(\mathbf{y}_s) k} \ket{k}_{\text{qpe}} \otimes \ket{\mathbf{y}_s}_{\text{qd}} \otimes \ket{0}_{\text{qb}}.
\end{equation}
$\tilde{F}_n$ include all normalizations. The QPE qubit states
\begin{equation}
\ket{\Phi(\mathbf{y}_s)} = \tilde{F}_n \sum_{k=0}^{2^l-1} e^{i2\pi \tilde{\phi}(\mathbf{y}_s) k} \ket{k}.
\end{equation}
Then
\begin{equation}
\ket{\psi_7} = \sum_{\mathbf{y}_s} \ket{\Phi(\mathbf{y}_s)}_{\text{qpe}} \otimes \ket{\mathbf{y}_s}_{\text{qd}} \otimes \ket{0}_{\text{qb}}.
\end{equation}
The full density matrix is
\begin{equation}
\rho = \sum_{\mathbf{y}_s, \mathbf{y}_{s'}} \ket{\Phi(\mathbf{y}_s)}\bra{\Phi(\mathbf{y}_{s'})}_{\text{qpe}} \otimes \ket{\mathbf{y}_s}\bra{\mathbf{y}_{s'}}_{\text{qd}} \otimes \ket{0}\bra{0}_{\text{qb}}.
\end{equation}
Tracing out QPE qubits and the qubit register:
\begin{equation}
\rho_{\text{qd}} = \Tr_{\text{qpe},\text{qb}}(\rho) = \sum_{\mathbf{y}_s, \mathbf{y}_{s'}} \braket{\Phi(\mathbf{y}_{s'})}{\Phi(\mathbf{y}_s)} \ket{\mathbf{y}_s}\bra{\mathbf{y}_{s'}}_{\text{qd}}.
\end{equation}
Let the distinct values of $\tilde{\phi}(\mathbf{y}_s)$ be $\{\phi_j\}_{j=1}^M$. Define
\begin{equation}
d_j = |\{ \mathbf{y}_s : \tilde{\phi}(\mathbf{y}_s) = \phi_j \}|, \quad \sum_{j=1}^M d_j = F_s,
\end{equation}
where $F_s$ is the total number of $\mathbf{y}_s$ in the sum. For QPE with $l$ qubits and well-separated phases,
\begin{equation}
\braket{\Phi(\mathbf{y}_{s'})}{\Phi(\mathbf{y}_s)} \approx \delta_{\tilde{\phi}(\mathbf{y}_s), \tilde{\phi}(\mathbf{y}_{s'})}.
\end{equation}
Thus
\begin{equation}
\rho_{\text{qd}} = \sum_{j=1}^M \frac{d_j}{F_s} \ket{\chi_j}\bra{\chi_j}, \quad \ket{\chi_j} = \frac{1}{\sqrt{d_j}} \sum_{\mathbf{y}_s : \tilde{\phi}(\mathbf{y}_s) = \phi_j} \ket{\mathbf{y}_s}.
\end{equation}
Eigenvalues: $d_j/S$ with multiplicity 1 for each $j$. The von Neumann entropy is
\begin{equation}
S(\rho_{\text{qd}}) = -\sum_{j=1}^M \frac{d_j}{F_s} \ln \frac{d_j}{F_s}.
\end{equation}
If all phases are distinct ($d_j = 1$ for all $j$, $M = F_s$):
\begin{equation}
S(\rho_{\text{qd}}) = \ln F_s.
\end{equation}
If all states share the same phase ($M = 1$, $d_1 = F_s$):
\begin{equation}
S(\rho_{\text{qd}}) = 0.
\end{equation}
\begin{figure*}[t]
\centerline{\includegraphics[width = 1\linewidth,trim={0cm 8cm 0cm 8cm},clip]{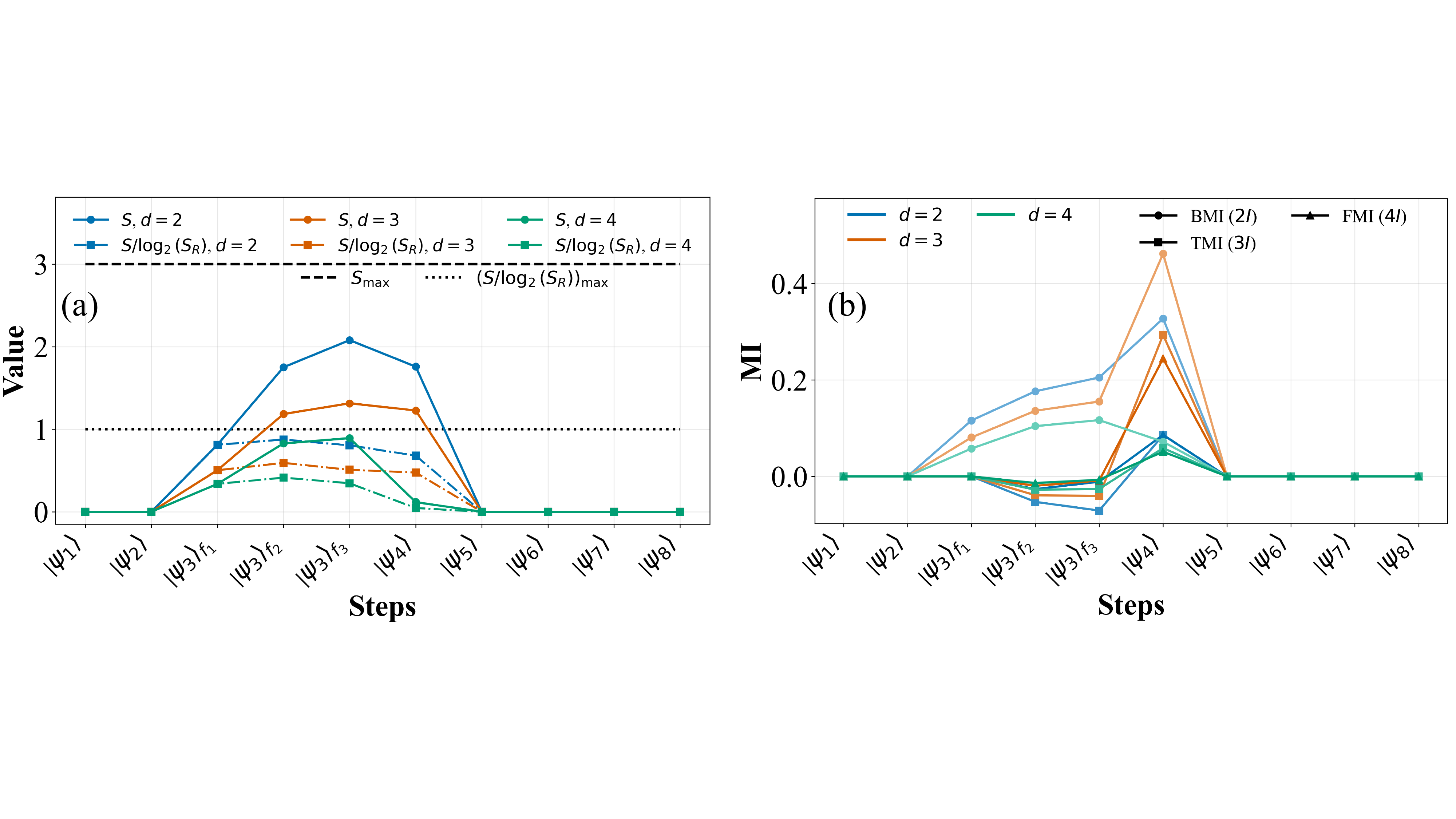}}
    \caption{Entanglement study for solving the example problem given by Eq.~\eqref{ExP1a} (a) von Neumann entropy $S$ across the qudit-qubit bipartition (solid lines) and the normalized ratio $S/\log_2(S_R)$ (dash-dotted lines) at each step of the algorithm, for qudit dimension $d=2$ (blue), $d=3$ (orange), and $d=4$ (green). The dashed black line marks the dimension-imposed maximum entropy $S_{\max}$, and the dotted black line marks $\log_2(S_R)$, the maximum entropy achievable given the Schmidt rank $S_R$ realized at that step. $S/\log_2(S_R)\to 1$ indicates a state that is maximally entangled for its own Schmidt rank, not relative to the full Hilbert-space dimension. (b) Bipartite (BMI, circles), tripartite (TMI, squares), and four-partite (FMI, triangles) mutual information as a function of algorithm step, for qudit dimension $d=2$ (blue), $d=3$ (orange), and $d=4$ (green), averaged over all subsystem partitions of the corresponding order.}
\label{entfig1}
\end{figure*}

\begin{figure*}[t]
\centerline{\includegraphics[width = 1\linewidth,trim={0cm 8cm 0cm 8cm},clip]{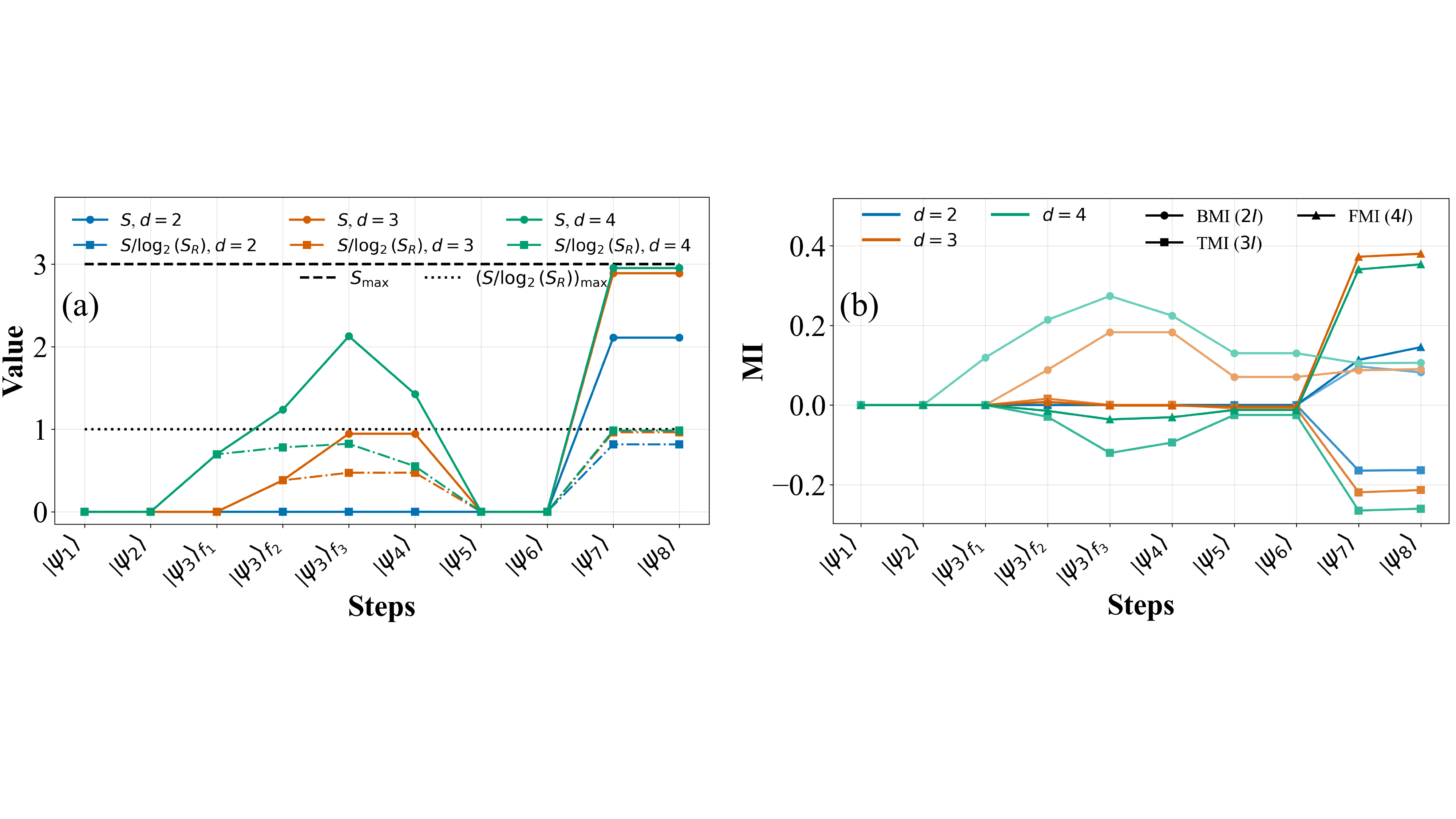}}
    \caption{Entanglement study for solving the example problem given by Eq.~\eqref{ExP1b} (a) von Neumann entropy $S$ across the qudit-qubit bipartition (solid lines) and the normalized ratio $S/\log_2(S_R)$ (dash-dotted lines) at each step of the algorithm, for qudit dimension $d=2$ (blue), $d=3$ (orange), and $d=4$ (green). The dashed black line marks the dimension-imposed maximum entropy $S_{\max}$, and the dotted black line marks $\log_2(S_R)$, the maximum entropy achievable given the Schmidt rank $S_R$ realized at that step. $S/\log_2(S_R)\to 1$ indicates a state that is maximally entangled for its own Schmidt rank, not relative to the full Hilbert-space dimension. (b) Bipartite (BMI, circles), tripartite (TMI, squares), and four-partite (FMI, triangles) mutual information as a function of algorithm step, for qudit dimension $d=2$ (blue), $d=3$ (orange), and $d=4$ (green), averaged over all subsystem partitions of the corresponding order.}
\label{entfig2}
\end{figure*}

\section{Entanglement generation for different problems}
\label{entanpeak}
The problem structure determines the generation of entanglement in the system. Consider two problems, one highly constrained (Problem 1) and the other with relaxed constraints (Problem 2) as defined below.
\begin{minipage}[t]{0.48\textwidth}
\begin{equation}
\begin{split}
&\text{PROBLEM 1}\\
C(\mathbf{x}) &= x_1x_2 + x_3x_4 + x_1x_3 + x_2x_4 + x_2 + x_3 \\
\text{To } &\text{be maximized under:} \\
C_1&: x_2 + x_3 \le 1 \\
C_2&: x_2x_1 + x_4 \le 1 \\
C_3&: x_1 + x_4 + x_3 \le 1 \\
x_i &\in \{0, 1, \dots, d\}.
\end{split}
\label{ExP1a}
\end{equation}
\end{minipage}
\hfill
\begin{minipage}[t]{0.48\textwidth}
\begin{equation}
\begin{split}
&\text{PROBLEM 2}\\
C(\mathbf{x}) &= x_1x_2 + x_3x_4 + x_1x_3 + x_2x_4 + x_2 + x_3 \\
\text{To } &\text{be maximized under:} \\
C_1&: x_2 + x_3 \le 5 \\
C_2&: x_2x_1 + x_4 \le 5 \\
C_3&: x_1 + x_4 + x_3 \le 5 \\
x_i &\in \{0, 1, \dots, d\}.
\end{split}
\label{ExP1b}
\end{equation}
\end{minipage}
\\
\\ \\
The Figs.~\ref{entfig1}~and~\ref{entfig2} show the entanglement for both the problems, respectively. For Problem 1, stage I contributes mainly while for Problem 2, it is the stage II that solves the problem. Similar behavior is shown in panel (b) of both the figures for different mutual information.

\begin{figure*}[t]
\centerline{\includegraphics[width = 1\linewidth,trim={0cm 6.5cm 0cm 7cm},clip]{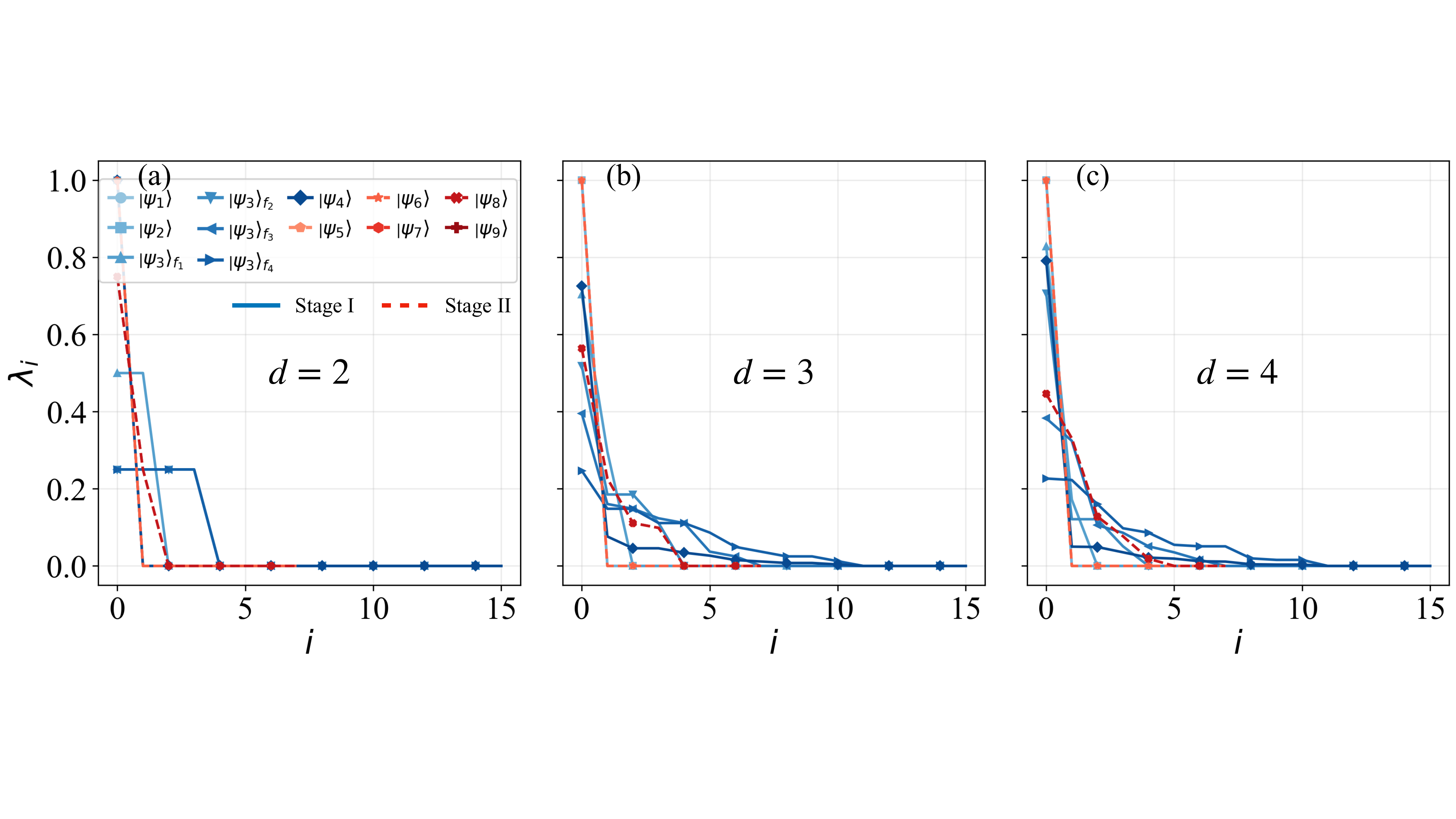}}
    \caption{Entanglement spectrum is shown along with the eigenvalues $\{\lambda_i\}$ of the reduced qudit density matrix $\rho_D$, sorted in decreasing order at each step of the algorithm, for qudit dimension $d=2$ [(a)], $d=3$ [(b)], and $d=4$ [(c)]. Solid curves are Stage-I states ($|\psi_1\rangle$-$|\psi_4\rangle$) and dashed curves are Stage-II states ($|\psi_5\rangle$--$|\psi_8\rangle$); $|\psi_3\rangle_{f_1}$--$|\psi_3\rangle_{f_4}$ label the state after each successive constraint unitary. These are for solving the problem given by Eq.~\eqref{ExP8}.}
\label{f3}
\end{figure*}

\section{Entanglement spectrum across the qudit-qubit cut}
\label{spectrumm}
The entanglement in the system is examined by calculating the entanglement spectrum of the subsystem remaining after tracing out all the qubits. The reduced density matrix of the qudit register is obtained by tracing out the qubits, $\rho_D = \Tr_Q(\dyad{\psi})$. Let $\{\lambda_k\}$ be the eigenvalues of $\rho_D$ in the decreasing order and the entanglement spectrum is the set of the eigenvalues as $\{-\log \lambda_k\}$.  The von Neumann entropy $S = -\sum_k \lambda_k \log \lambda_k$ is a single quantity calculated from this spectrum, however, the distribution of $\lambda_k$ contains more information of the distribution of the entanglement. 
The gap between the eigenvalues determines whether a truncation of the system is possible, a flat spectrum leads to long range correlations in the system.

\noindent \textbf{Numerical results}: In Fig.~\ref{f3} the entanglement spectrum corresponding to the states obtained in the algorithm for solving the problem described by Eq.~\ref{ExP8} for varying $d$ are presented. The entanglement spectrum is calculated by sorting the eigenvalues of the reduced density matrix for each eigenstate at the qudit–qubit cut. The states $\ket{\psi_1},\ket{\psi_2},\ket{\psi_6},\ket{\psi_7}$ exhibit a single unit eigenvalue with all others vanishing, it is a signature of unentangled product states. As the Stage-I progresses, although all the states display a dominant eigenvalue accompanied by a small number of non-negligible eigenvalues, the spectrum gets flatter which indicates that the entanglement is generated in the system due to the constraint-unitary operators. During the Stage II, the spectrum of states $\ket{\psi_8},\ket{\psi_9}$ coincide as QFT does not affect the entanglement structure while also showing a plateau of near-degenerate eigenvalues. This provides evidence of a more uniformly distributed entanglement structure and thus indicates that the problem instance is comparatively harder after satisfying the constraints. The qualitative behavior of each state's spectrum is similar as the dimension increases, suggesting that the entanglement structure is a robust feature rather than a finite-size artifact.
The flat spectrum is the defining characteristic of a Haar-random-like state across this bipartition, which is another indication that the output is in a regime that is classically intractable.
\twocolumngrid

\bibliographystyle{unsrt}
\bibliography{bib_file.bib}

\end{document}